\documentclass[apj,twocolumn,twocolappendix,numberedappendix]{openjournal}
\usepackage{newtxtext,newtxmath,enumitem,multirow,ctable,verbatim}
\usepackage[T1]{fontenc}
\usepackage[dvipsnames]{xcolor}
\usepackage[breaklinks,colorlinks,citecolor=blue,urlcolor=blue]{hyperref}

\newcommand{\paperone}{Paper {\small I}}
\newcommand{\papertwo}{Paper {\small II}}

\newcommand{\orcidauthor}[3]{\author{\href{http://orcid.org/#1}{#2$^{#3}$}}}

\shorttitle{CR-IC In Cool Cores}
\shortauthors{Hopkins et al.}

\begin{document}

\title{\vspace{-0.8cm}Cosmic Rays Masquerading as Cool Cores: \\
An Inverse-Compton Origin for Cool Core Cluster Emission\vspace{-1.5cm}}

\orcidauthor{0000-0003-3729-1684}{Philip F. Hopkins}{1*}
\orcidauthor{0000-0001-9185-5044}{Eliot Quataert}{2}
\orcidauthor{0000-0002-1616-5649}{Emily M. Silich}{1}
\orcidauthor{0000-0002-8213-3784}{Jack Sayers}{1}
\orcidauthor{0000-0002-7484-2695}{Sam B. Ponnada}{1}
\orcidauthor{0000-0002-1159-4882}{Isabel S. Sands}{1}
\affiliation{$^{1}$Division of Physics, Mathematics, and Astronomy, California Institute of Technology, Pasadena, CA 91125, USA}
\affiliation{$^{2}$Department of Astrophysical Sciences, Princeton University, Princeton, NJ 08544, USA}

\thanks{$^*$E-mail: \href{mailto:phopkins@caltech.edu}{phopkins@caltech.edu}},

\begin{abstract}
X-ray bright cool-core (CC) clusters ubiquitously contain luminous radio sources accelerating cosmic ray (CR) leptons at prodigious rates. 
Near the acceleration region, high-energy leptons produce synchrotron (mini)halos and sometimes observable $\gamma$-rays, but these leptons have short lifetimes and so cannot propagate far from sources without some rejuvenation. 
However, low-energy ($\sim 0.1-1\,$GeV) CRs should survive for $\gtrsim$\,Gyr, potentially reaching $\sim 100\,$kpc 
before losing most of their energy via inverse-Compton (IC) scattering of CMB photons to $\sim$\,keV X-ray energies, with remarkably thermal X-ray spectra.  In groups/clusters, this will appear similar to relatively ``cool'' gas in cluster cores, i.e.\ CCs. 
In lower-mass (e.g.\ Milky Way/M31) halos, analogous CR IC emission will appear as hot (super-virial) gas at outer CGM radii, explaining recent diffuse X-ray observations. 
We show that for plausible (radio/$\gamma$-ray observed) lepton injection rates, the CR-IC emission could contribute significantly to the X-ray surface brightness (SB) in CCs, implying that CC gas densities may have been significantly overestimated and alleviating 
the cooling flow problem.  A significant IC contribution to the diffuse X-ray emission in CC clusters also explains the tight correlation between the X-ray ``cooling luminosity'' and AGN/cavity/jet power even absent significant feedback heating, because the apparent CC emission is itself driven by the radio source.  
Comparing observed Sunyaev Zeldovich-to-X-ray inferred pressures at $\ll100$\,kpc in CCs represents a clean test of this scenario, and existing data appears to favor significant CR-IC.  A significant IC contribution also implies that X-ray inferred gas-phase metallicities have been underestimated in CCs, potentially explaining the discrepancy between X-ray (sub-Solar) and optical/UV (super-Solar) observed metallicities in the central $\sim 10\,$kpc of nearby CCs.  We also discuss our model's connection to observations of multiphase gas in clusters.
\end{abstract}


\maketitle

\section{Introduction}
\label{sec:intro}

Active galactic nuclei (AGN) and supernovae (SNe) are well-known to be sources of relativistic cosmic rays (CRs), including leptons and hadrons with most of their energy at $\sim 0.1-1\,$GeV \citep[e.g.][]{owen:2023.cr.review.galaxies.feedback,ruszkowski.pfrommer:cr.review.broad.cr.physics}. 
CR leptons from AGN are observed as radio (and sometimes $\gamma$-ray) sources in essentially every X-ray luminous ``strong cool core'' (SCC) cluster \citep{mcnamara:2007.agn.cooling.flow.review.cavity.jet.power.vs.xray.luminosity.scalings.emph.compilation,liu:2024.strong.lradio.lxray.connection.luminous.cool.cores.strong.central.xray.only.when.strong.radio.source}. High-energy leptons at $\gg 10\,$GeV produce most of the observed radio and $\gamma$-ray emission, but have short lifetimes; they thus radiate most of their energy near their production sites in regions of dense gas and strong magnetic/radiation fields.  The same sources should, however, also produce a large population of much longer-lived $\sim 0.1-1\,$GeV leptons, whose lifetimes $\gtrsim$\,Gyr imply that they can diffuse/advect/stream out to $\sim 100\,$kpc before losing most of their energy to inverse Compton (IC) scattering of cosmic microwave background (CMB) photons. 

In \citet{hopkins:2025.crs.inverse.compton.cgm.explain.erosita.soft.xray.halos} -- hereafter \paperone -- we noted that these extended, ``ancient'' cosmic ray halos or ``ACRHs'' would produce $\sim$ keV X-rays via IC scattering of CMB photons, with surprisingly
thermal continuum-like spectra (because the `aged' CR population is peaked around a GeV).
This could potentially explain the otherwise puzzling extended soft X-ray emission observed by ROSAT and eROSITA in the outskirts (near virial radii) of relatively low-mass (Milky Way/M31, i.e.\ virial temperatures $\sim 0.04-0.06\,$keV and below) halos \citep{anderson:2013.rosat.extended.cgm.xray.halos,zhang:2024.hot.cgm.around.lstar.galaxies.xray.surface.brightness.profiles,zhang:2024.erosita.hot.cgm.around.lstar.galaxies.detected.and.scaling.relations,zhang:2025.erosita.extended.halos.luminosity.versus.mass.versus.sf.quenched}.
These observations are, by contrast, in tension with thermal emission mechanisms given constraints from cosmology, galactic metal budgets, and other X-ray absorption constraints (see \paperone\ and \citealt{yao:2010.chandra.upper.limits.warm.hot.cgm.gas.in.mw.mass.halos,tumlinson:2017.cgm.review,ponti:2023.erosita.supervirial.gas.close.to.galaxy.cgm.low.metal.and.low.density,lau:2024.erosita.profiles.require.cosmological.constraints.violations,shreeram:2025.want.to.fit.erosita.w.illustris.have.to.change.halo.masses.and.renormalize.satellites.and.assume.central.psf.not.subtracted.and.psf.wrong.and.2x.count.satellite.lum.w.central.and.assume.all.at.third.solar.metal,vladutescu:2025.magneticum.erosita.profiles.xrb.contributions}). 
As also noted in \paperone, for massive groups and clusters, CR-IC emission cannot plausibly explain most of the diffuse X-ray emission out to the virial radius, which is dominated by thermal emission.
However, given the bright central radio sources observed in CCs, there could be appreciable CR-IC emission in cluster cores (within $\sim 100\,$kpc of those sources).  In this paper we expand on this idea, and show that it leads to a number of surprising conclusions that may call for fundamental revisions to our interpretation of the long-standing cooling flow problem in cool-core clusters.

Using simple analytic models (\S~\ref{sec:model}) for CR injection and CR transport (\S~\ref{sec:basic}), we illustrate how ACRHs produce remarkably thermal continuum-like X-ray spectra without strong radio/$\gamma$-ray counterparts (\S~\ref{sec:spectra}), and show how both the ``effective temperature'' (in X-rays) and characteristic radii scale with halo mass (\S~\ref{sec:mass}), and how the CR-IC emission properties should scale with cluster-centric radius in massive clusters (\S~\ref{sec:profiles}). 
We show that CR-IC appears as ``hot'' gas in the outskirts of low-mass halos, but ``cool'' gas in the cores of high-mass halos, producing surface brightness and X-ray inferred profiles strikingly similar to SCCs even when no true SCC is present. 
In \S~\ref{sec:cartoon} we discuss how this could imply fundamental revisions to our interpretation of the cooling flow problem in CCs.
\S~\ref{sec:obs} presents specific observational tests of this scenario with Sunyaev-Zeldovich (\S~\ref{sec:sz}) or high spatial-and-spectral-resolution X-ray abundances (\S~\ref{sec:metal}). 
We conclude in \S~\ref{sec:conclude}.

We  stress upfront that it has been suggested many times before that CR-IC could contribute to observed emission in some clusters \citep{raphaeli:1979.cluster.xray.emission.cr.ic.vs.thermal,sarazin:1988.book.cluster.xray.emission,sarazin:1999.cr.electron.emission.cluster.centers.xrays,hwang:1997.cluster.center.emission.in.euv.from.cr.ic}; there is indeed evidence for this is in many radio ``mini-halos'' \citep{gitti:2002.perseus.minihalo.xray.inverse.compton,gitti:2004.minihalo.abell.2626.reaccel,bonamente:2007.abell.3112.clear.xray.inverse.compton.required.luminosity.fits.models.gamma.rays.too,murgia:2010.ophiuchus.cluster.minihalo.xray.inverse.compton,bartels:2015.radio.inverse.compton.cluster.minihalo.prospects,gitti:2016.radio.minihalos.coolcore.clusters.candidates.review}.   However, these previous considerations of IC emission by CRs have  focused on the younger, more radio-bright CRs, which produce harder X-ray emission.   The focus in this paper is on more spatially extended softer X-ray emission produced by lower energy CR electrons.

\begin{figure}
	\includegraphics[width=1.02\columnwidth]{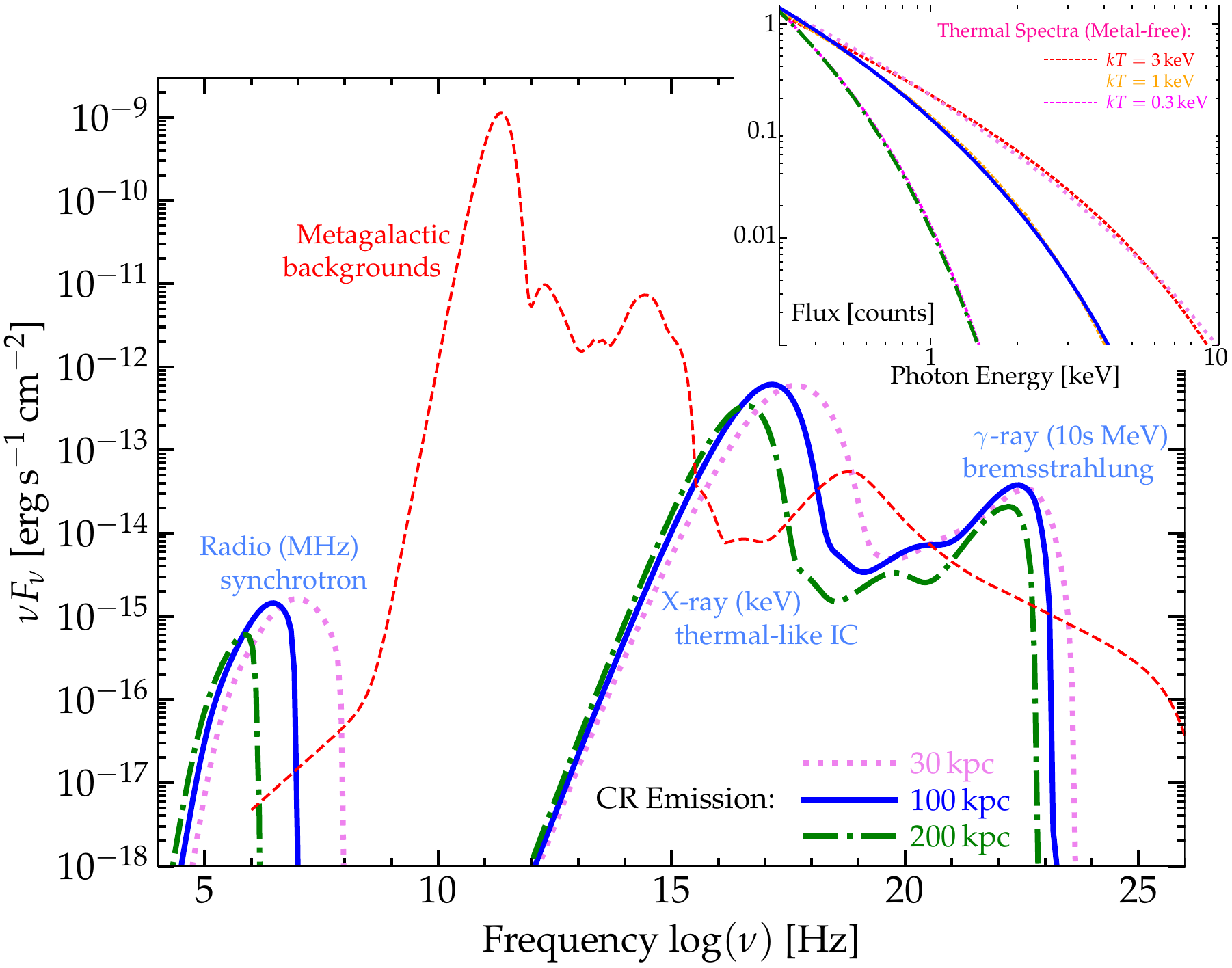}
	\caption{CR emission flux for an extended ancient CR halo at three different radii (\textit{labeled}), for the simple model in \S~\ref{sec:model} of CRs with a constant diffusivity+streaming speed in a galaxy cluster of $M_{\rm vir}=10^{15}\,M_{\odot}$ at $z=0.05$, with steady-state injection rate $\dot{E}_{\rm cr,\,\ell} \sim 10^{44}\,{\rm erg\,s^{-1}}$ ($\dot{E}_{43}=10$) from a central source (e.g.\ bright radio galaxy). 
	We compare metagalactic background light \citep{cooray:2016.extragalactic.background.light.compilations.review,khaire:2019.extragalactic.background.light.spectra,tompkins:2023.cosmic.radio.background} in a $100\,$kpc aperture, and show (\textit{inset}) the flux (arbitrarily normalized in ${\rm counts\,s^{-1}\,keV^{-1}}$) at $0.3-10\,$keV X-rays, compared to thermal spectra of primordial gas.
	Most of the CR emission comes from CR-IC scattering CMB photons, producing a thermal continuum-like ($\sim$\,keV) soft X-ray peak. 
	The secondary peaks with low-surface brightness at $10-100\,$MeV $\gamma$-rays (from relativistic bremsstrahlung) and $\sim 0.5-10\,$MHz radio (from synchrotron) are unobservable at present, though the radio is the natural evolution of known ultra-steep and GHz sources on smaller ($\lesssim 10\,$kpc) scales.
	\label{fig:spectrum}}
\end{figure}

\section{What Does an ACRH Look Like?}
\label{sec:model}

In a companion paper, \citet{hopkins:2025.cr.ic.clusters.detailedobs} (henceforth \papertwo), we present detailed calculations and scalings for a variety of properties of CR-IC emission in clusters, and comparison with dozens of different observations. Here, for brevity and clarity, we describe more qualitatively the scenario at hand, emphasizing its key points.  We refer to \papertwo\ for the detailed model scalings and equations solved.

\subsection{``Initial Conditions'' and Basic Background}
\label{sec:basic}

Consider a galaxy or AGN with escaping CR energy per unit time in leptons $\dot{E}_{\rm cr,\,\ell} \sim 10^{43}\,{\rm erg\,s^{-1}}\,\dot{E}_{43}$ (with some  distribution of CR kinetic energies $E_{\rm cr} \equiv E_{\rm cr,\,GeV}\,{\rm GeV}$). In the CGM, the spherically-averaged profile of the lepton energy density $e_{\rm cr,\,\ell}(R)$ at radius $R$ can then be represented as $e_{\rm cr,\,\ell} \sim {f_{\rm loss} \dot{E}_{\rm cr,\ell}}/{4\pi\,v_{\rm st,\,eff} R^{2}} \sim 0.5 f_{\rm loss} {\rm eV\,cm^{-3}}\,({\dot{E}_{43}}/{(R/100\,{\rm kpc})^{2}\,(v_{\rm st,\,eff}/100\,{\rm km\,s^{-1}})})$, where $f_{\rm loss}$ accounts for CR radiative losses, and $v_{\rm st,\,eff}$ is some effective, isotropically-averaged bulk streaming speed of CRs (including advection, microphysical streaming, diffusion, etc., and allowed to depend on position/time/$E_{\rm cr}$). Motivated by standard models for acceleration in jets/radio galaxies and modeling of compact radio and $\gamma$-ray sources in clusters, we assume primarily leptonic injection (distinct from e.g.\ supernovae-accelerated CRs in \paperone), but consider varying lepton-to-hadron ratios (largely unimportant for our conclusions here) in \papertwo. At a time $\Delta t \equiv \Delta t_{\rm Gyr}\,{\rm Gyr}$ since their ``injection'' (here referring to their escape from the central galaxy or small-scale acceleration region) into the CGM, they will have propagated some average distance $R \sim v_{\rm st,\,eff} \Delta t \sim 100\,{\rm kpc}\,(v_{\rm st,\,eff}/100\,{\rm km\,s^{-1}})\,\Delta t_{\rm Gyr}$. But as they propagate, they will also lose energy to inverse Compton+synchrotron, Coulomb+ionization interactions, and bremsstrahlung \citep{blumenthal:1970.cr.loss.processes.leptons.dilute.gases}. Comparing loss timescales from these different processes, for any reasonable CGM densities $n \sim n_{-3}\,10^{-3}\,{\rm cm^{-3}}$ and magnetic field strengths $B \sim B_{\rm \mu G} {\rm \mu G}$, the dominant loss channel at $\gtrsim 0.1\,$GeV CR energies is IC scattering of CMB photons, with loss timescale $t_{\rm loss,\,IC} \sim 1.2\,{\rm Gyr}\,(1+z)^{-4}\,E_{\rm cr,\,GeV}^{-1}$ \citep{blumenthal:1970.cr.loss.processes.leptons.dilute.gases}. So high-energy ($\gg$\,GeV) CRs lose their energy rapidly, suppressing the spectrum above an energy $E_{\rm cr,\,GeV} \gtrsim 1.3\,(1+z)^{-4}\,(v_{\rm st,\,eff}/100\,{\rm km\,s^{-1}})\,(R/100\,{\rm kpc})^{-1}$. 

Initially, in the ``injection zone,'' the source therefore resembles some standard CR source. For Milky Way (MW)-like galaxies, the injection zone is the ISM, and represents the collection of all SNe, pulsars, etc., together with the central AGN.  The loss and transport processes in that ISM lead to a CR spectrum like that observed in the local ISM (LISM) -- aka the far outskirts of the ISM -- before CRs escape (with $\dot{E}_{43} \sim 10^{-3}$) into the CGM (\citealt{dimauro:2023.cr.diff.constraints.updated.galprop.very.similar.our.models.but.lots.of.interp.re.selfconfinement.that.doesnt.mathematically.work,silver:2024.cr.propagation.low.energies.new.data,delatorre.luque:2024.gas.models.of.galaxy.key.for.scale.height.but.need.halo.for.crs}).    

In massive galaxy clusters, the ``injection zone'' is the near-vicinity of the central AGN/blazar/jet/radio lobes and/or associated X-ray cavities. Essentially all SCC clusters contain luminous radio sources accelerating CRs at a prodigious rate -- quantitatively, this means $\dot{E}_{43} \sim 0.1-100$, with typical injection-zone synchrotron spectral index $\alpha \sim 1$, somewhat shallower than the LISM lepton spectrum and (usually inferred) much higher lepton-to-hadron ratio; \citealt{osullivan:2011.radio.jet.power.radio.emission.and.xray.cooling.luminosities,ignesti:2020.cluster.minihalo.leptonic.edot.1e44.1e46.signatures.radio.and.xrays,keenan:2021.jet.leptonic.power.1e41to1e45.easily.produced.from.modest.agn.bursts.or.steady.jets,foschini:2024.blazar.agn.jet.power.favor.leptonic.large.power.energy.much.more.than.kinetic.lobe.cavity.power,whittingham:2024.cluster.Bfield.synch.measurements.biased.to.strongest.B.subregions}). These central sources can be detected across a wide range of wavelengths as blazars/AGN: GHz radio, hard X-ray (power-law-like IC), $\gamma$-rays \citep{wik:2011.swift.ic.hardx.upper.limits.clusters.b.lower.point1microg,ackermann:2014.cosmic.ray.fermi.gamma.ray.upper.limits.galaxy.clusters.data.not.as.model.dependent,cova:2019.cluster.ic.upper.limits.B.lower.limits,mirakhor:2022.ic.cluster.detection.B.0pt1microG}. 
The termination of the active jets can be associated with radio-bright, X-ray dim X-ray cavities (believed to be inflated by said CRs), which define an extended effective injection zone.
But assuming CRs exit these regions, after $\Delta t \gtrsim 10^{7}\,{\rm yr}$ ($\Delta t_{\rm Gyr} \gtrsim 0.01$, $R \gtrsim $\,kpc), \textit{just} including losses from the CMB (i.e.\ ignoring stronger $B$ or radiation fields in the injection zone that give stronger synchrotron/IC losses), losses cut off the higher-energy CR spectrum, so the diffuse CR emission at $R \gg$\,kpc is generally not detectable as a GHz radio source (and the hard X-ray and $\gamma$-ray components also disappear), but only as an ultra-steep spectrum $\sim 100\,$MHz radio halo \citep{savini:2018.lofar.ultrasteep.radio.emission.surrounding.minihalo.larger.radii.steepening.as.expected.in.coolcore,cuciti:2021.diffuse.cluster.radio.halo.fluxes.brightness.lofreq.gmrt.steeper.slopes.larger,biave:2021.lofar.ultrasteep.low.freq.radio.cluster.halo.rejuvenated.recently,ignesti:2022.lofar.zdrop.cluster.central.ultrasteep.radio.relic.lofar.losing.energy.outside.of.center,edler:2022.abell1033.cluster.case.study.lofar.decaying.crs.super.steep.radio.rejuvenated.at.special.location.older.at.larger.r.as.expected,pasini:2024.lofar.low.freq.radio.relic.detection.tend.to.extremely.steep.slopes}. The CR spectrum continues to soften as CRs propagate further from the cluster center (and likely, $B$ weakens further from the cluster center); by $\Delta t \gg 10^{8}$\,yr or $R \gg 10\,$kpc, the radio emission is undetectable at $\gtrsim$\,MHz frequencies, and the system is an ACRH.

To provide some intuition for when CR-IC from the ACRH can be important, it is instructive to compare the keV emissivity from CR-IC of CMB photons to the thermal free-free emissivity.  The result is
\begin{equation}
{\left. \frac{\epsilon_{IC}}{\epsilon_{ff}} \right|}_{\rm keV} \sim 100\,(1+z)^{4} \, \frac{p_{\rm CR, \ell}}{p_{\rm th}} \left(\frac{T}{10^8 \, {\rm K}}\right)^{3/2} \left(\frac{n}{0.01 \, {\rm cm^{-3}}}\right)^{-1}
\label{eq:emissivity_ratios}
\end{equation}
Equation \ref{eq:emissivity_ratios} shows that for plausible NCC temperatures and gas densities CR-IC can readily outshine thermal free emission in the soft X-rays even with a CR lepton pressure that is just a modest fraction of the thermal gas pressure.

Note we are specifically interested in the CRs accelerated by the central sources in these galaxies, as these will be strongly concentrated in a relatively steep profile around those sources. These are quite distinct from the volume-filling populations out to $\sim$\,Mpc radii associated with shock acceleration and potentially turbulent re-acceleration in famous giant radio halos (e.g.\ Coma). Those have very different spectra and physical origins, and much lower-surface-brightness/extended/non-concentrated radial profiles, and as such are not predicted by the modeling here to contribute significantly to X-rays via CR-IC.

\subsection{Resulting Spectra}
\label{sec:spectra}

Fig.~\ref{fig:spectrum} shows the results of a more detailed calculation for a cluster with $M_{500}=10^{15}\,M_{\odot}$ at $z=0.05$; these results, and what follows, assume pure lepton injection, i.e., we neglect the possible proton contribution.    Beginning from either a pure-power-law injection spectrum in momentum-space ($dN_{\rm cr}/dp \propto p^{-4.2}$, i.e.\ $\delta = 2.2$) or LISM-like CR spectrum for the ``injection zone'' (assumed leptonic; \citealt{bottcher:2013.blazar.modeling.almost.all.blazars.better.fit.by.leptonic.cr.models.not.hadronic,blandford:2019.agn.jets.review,cerruti:2020.agn.jet.leptonic.hadronic.review}), we propagate a population of CRs assuming a steady-state injection-zone spectrum with said spectral shape and $\dot{E}_{43}=10$. 
Note that this is specifically chosen to be the median \textit{observationally-inferred} CR lepton injection rate (and spectrum) from e.g.\ the correlation between central radio galaxy compact source luminosity and X-ray apparent cooling luminosity \citep[see][]{birzan:2004.radio.cavity.jet.power.vs.xray.power.correlation,osullivan:2011.radio.jet.power.radio.emission.and.xray.cooling.luminosities,sun:2009.every.radio.agn.has.xray.cc.strong.lradio.lxray.connection.clusters.all.xray.agns.also.radio.agns,mittal:2009.radio.cluster.properties.vs.xray.strong.radio.fraction.increases.with.coolcore.strength.all.scc.strong.radio.correlation.with.mdot.lx,liu:2024.strong.lradio.lxray.connection.luminous.cool.cores.strong.central.xray.only.when.strong.radio.source}, so we are ensured that ``sufficient CRs exist'' for the models here and these do not (by construction) violate any limits on central radio luminosities. We solve for the CR radial distribution by solving the steady-state diffusion-streaming-advection (standard Fokker-Planck-like) equation for CR transport \citep{zank:2014.book,hopkins:m1.cr.closure,thomas:2021.compare.cr.closures.from.prev.papers} assuming a constant effective (isotropically-averaged) diffusivity $\kappa_{\rm iso} \sim 10^{29}\,{\rm cm^{2}\,s^{-1}}$ at GV (fit to standard Solar-system CR data with GALPROP in \citealt{korsmeier:2021.light.element.requires.halo.but.upper.limit.unconfined,dimauro:2023.cr.diff.constraints.updated.galprop.very.similar.our.models.but.lots.of.interp.re.selfconfinement.that.doesnt.mathematically.work,silver:2024.cr.propagation.low.energies.new.data,tovar:2024.inhomogeneous.diffusion.cr.spectra}, and scaling with rigidity as therein)  and $v_{\rm st,\,eff} \sim 100\,{\rm km\,s^{-1}}$ (motivated by CGM observations in \citealt{karwin:2019.fermi.m31.outer.halo.detection,butsky:2022.cr.kappa.lower.limits.cgm,hopkins:2025.crs.inverse.compton.cgm.explain.erosita.soft.xray.halos}), accounting for energy-dependent Coulomb, ionization, bremsstrahlung, synchrotron, and inverse Compton losses  (following \citealt{blumenthal:1970.cr.loss.processes.leptons.dilute.gases,1972Phy....60..145G,ginzburg:1979.book,rybicki.lightman:1979.book}).  We take the background gas density+temperature profile from the ``universal'' fits in \citet{ghirardini:2019.cluster.profiles.compilation.universal.fits} (scaled to $T_{\rm vir}$ and $R_{500}$, such that the universal baryon fraction is found within $R_{\rm vir}$), and simply fix $\beta \equiv P_{\rm therm}/P_{\rm mag} = n k T/(B^{2}/8\pi) \sim 100$ (relatively strong fields, in clusters, but it does not change our conclusions significantly if we just take a constant $B \lesssim 3\,{\rm \mu G}$ for the diffuse $\sim 100\,$kpc gas; \citealt{bohringer_cluster_RMs_2016}). 

We see three spectral peaks: $\sim 1-$few\,MHz radio synchrotron, $\sim 10-100\,$MeV $\gamma$-ray bremsstrahlung, and $\sim$\,keV IC. The radio and $\gamma$-ray peaks\footnote{Note that the claim that $\gamma$-rays imply $\lesssim 1\%$ of cluster pressure in CRs in e.g.\ \citet{ackermann:2014.cosmic.ray.fermi.gamma.ray.upper.limits.galaxy.clusters.data.not.as.model.dependent} only apply, as stated therein, if several assumptions are true including (1) all CRs are hadronic (there is no leptonic constraint); (2) CRs transport is strictly adiabatic and CRs do not diffuse/stream; (3) CRs have very hard power-law spectra to $\gg$\,GeV; and (4) the CR-to-gas pressure is uniform to $R_{500}$ (where the constraint is measured), rather than the CRs being concentrated in the CC. Clearly none of these apply to the models here, and as a result the predicted $\gamma$-ray luminosities are orders-of-magnitude lower than the Fermi upper limits quoted therein.}  are modeled in more detail and compared directly to observations (for a range of model assumptions beyond those here) in \papertwo, but we ignore them for now in this initial manuscript as they are undetectable (both because of their low frequency -- $<10\,$MHz and $<100$\,MeV, respectively -- and their extremely low surface brightness/flux -- $\ll {\rm mJy\,arcsec^{-2}}$ and $\ll 10^{-12}\,{\rm photons\,s\,cm^{-2}}$, respectively) in any current instrument, even if we boost the luminosity to $\dot{E}_{43}=100$ 
(though the $\lesssim 100\,$MHz radio within $\ll 100\,$kpc may be detectable in the brightest radio halos e.g.\ M87/Virgo and 3C 84/Perseus, and in future work we show these appear to be consistent with the same leptons needed to explain their core soft X-ray observations via CR-IC). Outside these three peaks (e.g.\ IR/optical/UV/EUV) the emission is orders-of-magnitude below backgrounds and other sources (e.g.\ intra-cluster light; see \papertwo).

But the soft X-ray peak is easily detectable. As shown in the inset in Figure \ref{fig:spectrum} (see also \paperone), because the X-rays are produced by IC scattering of a perfect blackbody (the CMB), and the higher-energy CR spectrum is truncated by losses, the X-ray spectrum is remarkably thermal-continuum-like, peaked around $\sim$\,keV. As shown in more detail in \papertwo, this sort of spectral shape would clearly not be detected by traditional searches for IC emission in hard X-rays (which search for hard power-law components), and it is obviously very different from a pure power-law as often assumed. 
The peak comes simply from the fact that most of the CR lepton energy is in $\sim 0.1-1$\,GeV CRs, which up-scatter a typical CMB photon to $\gamma^{2}\,k\,T_{\rm cmb} \sim $\,keV (without an anomalous harder or softer component). 
As we note below, this suggests a very simple empirical modeling approach: CR-IC X-rays from a multi-age CR distribution can be approximated in popular codes like XSPEC and APEC as a sum of primordial gas thermal contributions, together with the usual (metal-enriched) multi-phase thermal gas, where the effective temperature of the CR-IC components relates to their age. 

The $\sim$\,keV soft X-ray emission eventually cuts off at $\Delta t_{\rm Gyr} \gg 1$, i.e.\ $R \gg 100\,$kpc, as IC losses remove most of the energy from $\sim 0.1-1$\,GeV CRs (while Coulomb/ionization losses remove energy from $\ll 0.1$\,GeV CRs). So from scales of $\sim 1 - 100$\,kpc, CR-IC halos would appear similar to bright, thermal X-ray continuum emission.

Of course, the total spectrum observed will include many additional components beyond the direct CR emission continuum in Fig.~\ref{fig:spectrum}. In X-rays it will include line emission (both from the true thermal gas present, and from CR-excited emission) which can produce significantly different spectral properties. We briefly discuss this below in \S~\ref{sec:metal}, and in more detail in \papertwo, where we show the results of more realistic detailed X-ray spectral modeling of the combined (CR-IC+thermal) spectra from $\sim 0.1-10\,$keV, from both low-spectral-resolution (CCD) and high-spectral-resolution ($\lesssim 1\,$eV microcalorimeter) observations, allowing for the combined effects of continuum shape, line-ratio, and thermal broadening/linewidth variations. But we show these do not change our key conclusions here.

We also stress that we are not arguing that harder components of the spectrum cannot exist. Extended/diffuse, harder/shallower-spectrum radio or $\gamma$-ray or very hard X-ray emission could, of course, be present (in addition to the leptons we model here) from e.g.\ extended injection or acceleration or re-acceleration regions (e.g.\ strong shocks or jet termination regions), or other sources (e.g.\ AGN in the cluster). Their presence or absence has no impact on our conclusions here.

\begin{figure}
	\centering\includegraphics[width=0.99\columnwidth]{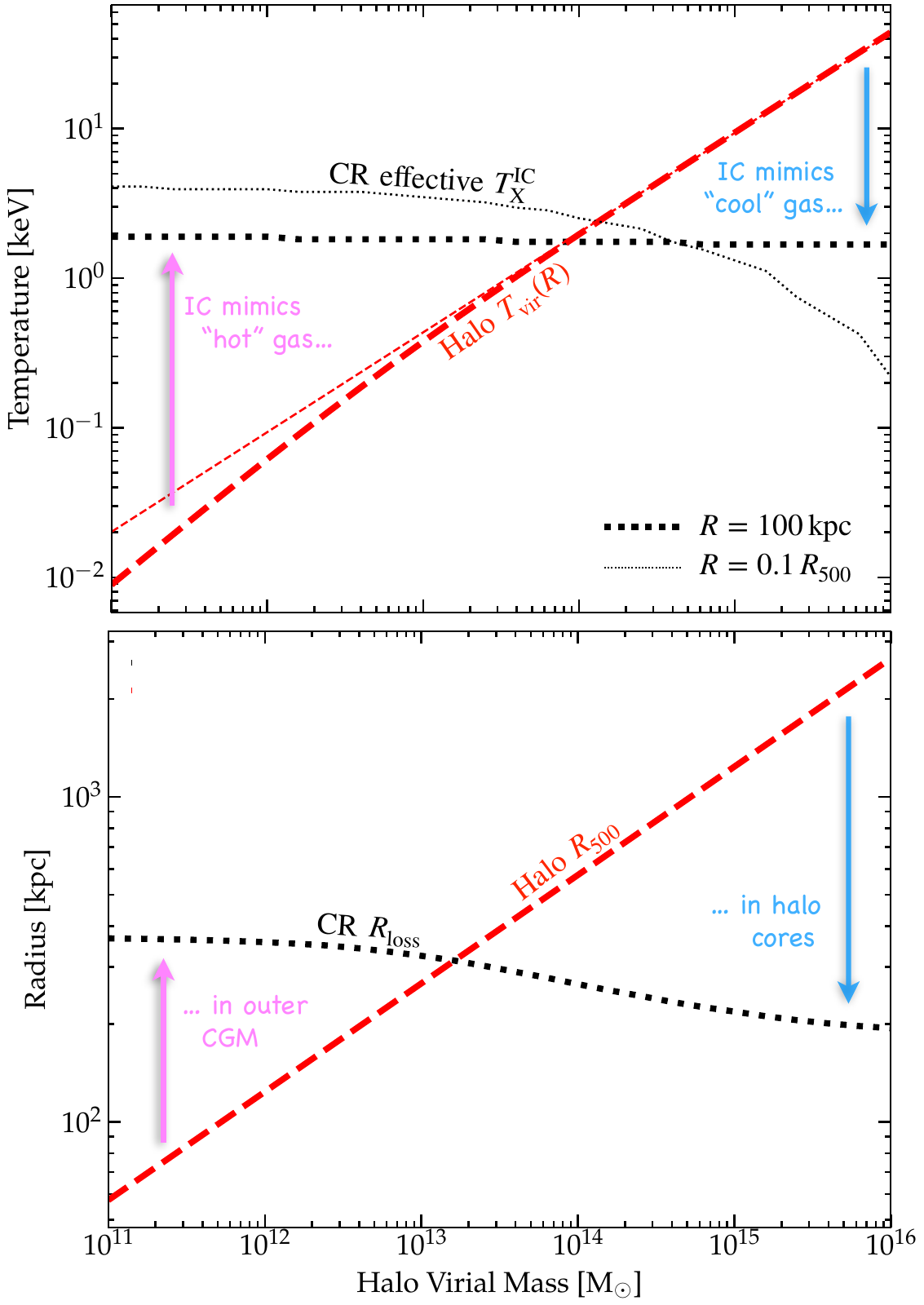} 
	\caption{Halo mass ($M_{\rm vir}$) dependence of the qualitative properties of the extended CR X-ray emission around galaxies. 
	For each halo we assume the self-similar model in \S~\ref{sec:model} (Fig.~\ref{fig:spectrum}).
	{\em Top:} CR emission ``effective'' (best-fit thermal from $0.1-10\,$keV) X-ray temperature $T_{\rm X}^{\rm IC}$, and actual gas virial temperature $T_{\rm vir}$, at either $0.1\,R_{500}$ or $100\,$kpc, versus $M_{\rm vir}$.
	{\em Bottom:} Radii where losses truncate the CR emission sharply (\S~\ref{sec:mass}; assuming a constant CR streaming speed), and $R_{500}$, versus $M_{\rm vir}$.
	Because of the relative scalings of these quantities, in intermediate/low-mass halos (Milky Way/M31, or $\lesssim 10^{13}\,M_{\odot}$), CR-IC will generically mimic ``hot'' (super-virial) gas in the \textit{outer} CGM around galaxies (near or beyond the virial radius).
	In high-mass halos (group/cluster, $\gtrsim 10^{14}\,M_{\odot}$), CR-IC will mimic ``cool'' (sub-virial) gas in the cluster \textit{cores}. 
	The crossover point ($\sim 10^{13}-10^{14}\,M_{\odot}$) depends on details of the profiles and redshift. 
	\label{fig:mhalo}}
\end{figure}

\subsection{Dependence on Halo Mass}
\label{sec:mass}

How would this thermal continuum-like CR-IC emission be interpreted, in halos of different masses?  In Fig.~\ref{fig:mhalo}, we show how (1) the effective temperature $T_{\rm eff,\,IC}$ (best-fit thermal X-ray temperature as Fig.~\ref{fig:spectrum}) of the CR-IC emission, and (2) effective size $R_{\rm loss}$ of the ACRH emission (taken for simplicity as the radius at which $f_{\rm loss}$ drops below $<0.1$) scale according to the model above, in halos of different masses. We compare these to the virial temperature $T_{\rm vir}$ and radii $R_{500}$ of the halos.  Note that for these calculations (and what follows) we assume for simplicity that the effective streaming speed is a constant $\sim 100 \, {\rm km \, s^{-1}}$ independent of halo mass, but the results are not sensitive to factor of few variations in $v_{\rm st, eff}$.

In low-mass systems, e.g.\ the MW and M31 and dwarf galaxies, $T_{\rm eff,\,IC} \gg T_{\rm vir}$ and $R_{\rm loss} \gtrsim R_{500}$. In other words, CR-IC X-ray emission will mimic extended, outer halo/CGM ``hot'' gas emission. This is the main point in \paperone, where we show that this simple model of CR IC X-ray emission provides a remarkably good fit to the $0.5-2$\,keV diffuse eROSITA X-ray halos observed around galaxies of these masses \citep{zhang:2024.hot.cgm.around.lstar.galaxies.xray.surface.brightness.profiles,zhang:2024.erosita.hot.cgm.around.lstar.galaxies.detected.and.scaling.relations}.
In high-mass systems, e.g.\ X-ray groups and clusters ($\gg 10^{14}\,M_{\odot}$), the opposite is true, $T_{\rm eff,\,IC} \lesssim T_{\rm vir}$ and (if $v_{\rm stream}$ does not increase rapidly with cluster mass) $R_{\rm loss} \ll R_{500}$. So in these systems, CR-IC will mimic ``cool'' gas emission from the halo core. But that is precisely what we see in cool-core (CC) groups/clusters in X-ray observations. So this raises the question, could CR-IC contribute significantly to observed CC X-ray emission?

\begin{figure*}
	\centering\includegraphics[width=0.99\textwidth]{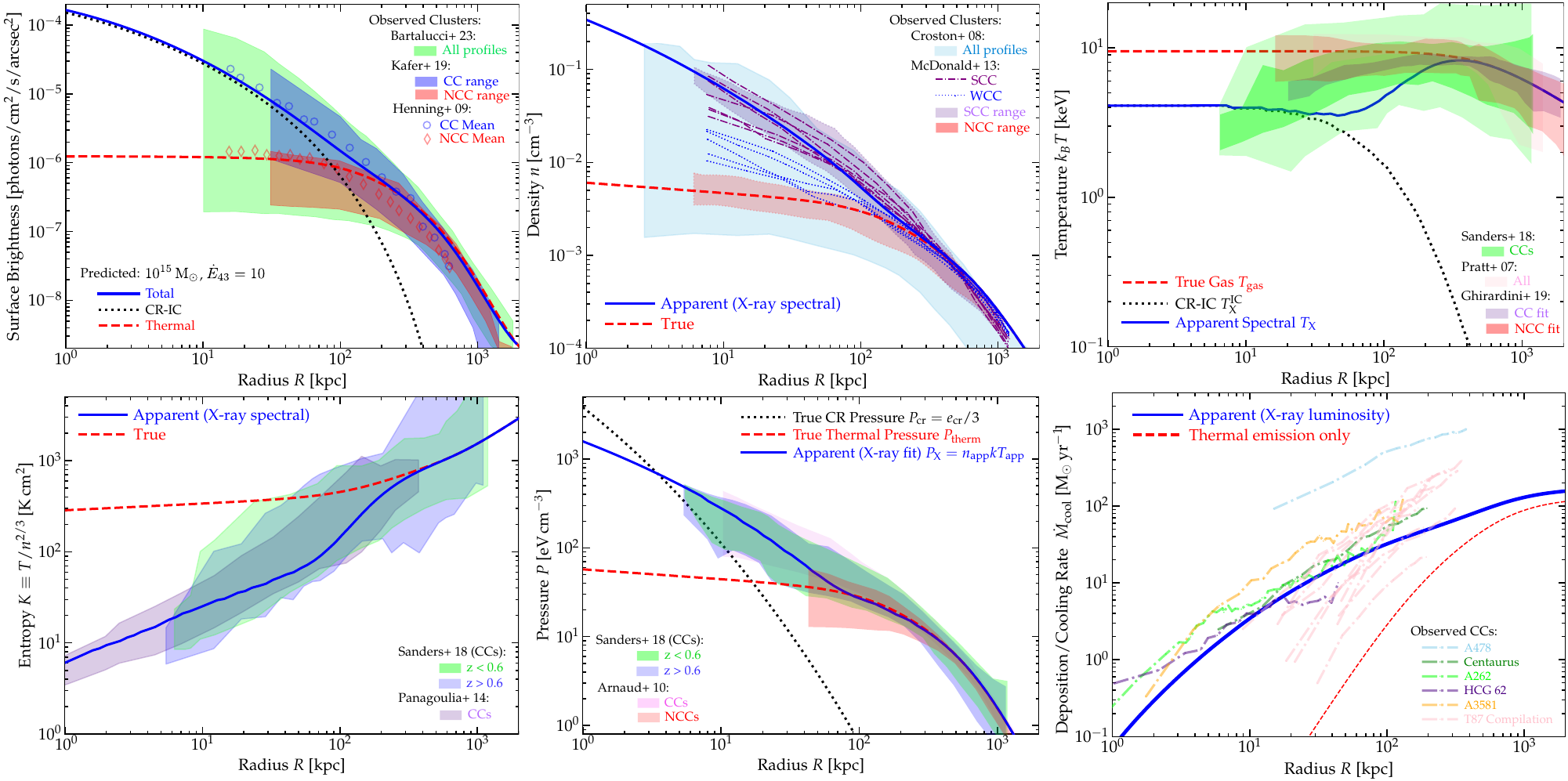} 
	\caption{Projected radial profiles of X-ray derived quantities for the same model of a galaxy cluster with a luminous CR injection source as Fig.~\ref{fig:spectrum}. The ``true'' gas properties are taken to follow typical non cool-core (NCC) clusters (\S~\ref{sec:profiles}), to illustrate an extremal case where all the ``cool core'' arises because of CR-IC. 
	We compare the projected surface brightness (versus 2D radius), gas density $n$, temperature $T$, entropy $K$, thermal pressure $P=n k_{B} T$, and ``mass deposition'' rate $\int_{0}^{r} (2\,\mu\,m_{p}/k T)\,dL_{\rm X,\,cool}$ versus (3D) radius $r$. 
	For each we show the ``true'' profile from the assumed NCC profile, and the estimated ``apparent'' profile accounting (very simply) for the contribution of CR-IC to the X-ray spectrum from $0.1-10$\,keV (\S~\ref{sec:profiles}).
	For a modest CR injection luminosity expected in typical bright radio galaxies, CR-IC could easily contribute much of the observed X-ray surface brightness at $\lesssim 100\,$kpc, leading to an over-estimate of density, pressure, and apparent mass deposition rate (cooling flow strength), and under-estimate of temperature and entropy, in the cluster core (when one assumes the emission is all thermal). The toy model profiles here are remarkably similar to strong cool cores observed, and the range of CC profiles observed can be reproduced by varying $\dot{E}_{\rm cr,\,\ell}$, despite there being (by construction) little or no ``true'' cool thermal core in the model.	
	\label{fig:profiles}}
\end{figure*}

\subsection{Radial Profiles in Clusters and Similarity to Cool Cores}
\label{sec:profiles}

To estimate what the effect of CR-IC would be in cluster centers, and whether this could mimic CCs, let us consider the most extreme possible case. Specifically, assume the gas properties follow a \textit{non}-cool-core (NCC) cluster profile, to mimic a plausible minimum contribution from thermal cooling emission.   In addition, we assume that there is a source of CRs with $\dot{E}_{43} \sim 10$, typical of many bright cluster radio galaxies. We then calculate the CR emission  as in \S~\ref{sec:spectra}, and calculate the thermal emission from the (assumed single-temperature) gas using APEC \citep{foster:2012.apec} to get a combined spectrum at each radius $R$.

In Fig.~\ref{fig:profiles}, we plot (1) the surface brightness (SB) profile, showing contributions from thermal and CR-IC; and both the ``true'' and ``apparent'' (2) density $n_{\rm gas}$ and (3) temperature $T$ profiles \textit{inferred} from the combined spectrum. For the latter, rather than model specific instrument response functions and fitting procedures (which can differ widely), we simply take the effective values to be the values of $n_{\rm gas}$ and $T_{\rm gas}$ that would give the same SB and $\sim 0.1-10$\,keV average spectral slope/hardness ratio of the spectrum (so this is an ``effective weighted temperature,'' defined akin to multi-temperature plasmas modeling X-ray observations in \citealt{mazzotta:2004.xray.temperature.measurement.modeling.and.caveats}). We then plot (4) the derived ``entropy'' $K\equiv T/n^{2/3}$, (5) pressure $P \equiv n\,k_{B} T$, and (6) the apparent ``mass deposition rate'' or ``mass cooling rate'' $\dot{M}_{\rm cool} = \int_{0}^{r} (2\mu m_{p}/5\,k_{B} T)\,{\rm d} L_{\rm X,\,cool}(r)$.

We immediately see (1) the predicted CR-IC can dominate the observed SB in the central $\lesssim 100\,$kpc, leading (2) to an inferred steeply-rising central  gas density (since SB scales $\propto n^{2}$), with (3) much more weakly-falling $T$ to about $\sim T_{\rm vir}/3$ (since the effect of $T$ on the spectral shape is weak, and $T_{\rm eff,\,IC}$ is not wildly different from $T_{\rm vir}$). This leads to (4) an apparently falling central entropy $K$, and (5) over-estimate of the pressure $P$. This (6) results in a large apparent $\dot{M}_{\rm cool}(r)$, which rises linearly with radius until saturating near the apparent ``cooling radius.''
We overplot observed profiles of CCs \citep{thomas:1987.cluster.mass.deposition.rates.coolingflows.compilation,white:1994.a478.cluster.cooling.flow.mass.deposition.profile.very.bright.cc,pratt:2007.cluster.temperature.profiles,croston:2008.cluster.density.profiles,henning:2009.xray.cluster.sb.profiles.obs.vs.sims,sanders:2010.cc.profiles.highres.deposition.rate.scalings,arnaud:2010.cluster.pressure.profile.fits,mcdonald:2013.cluster.gas.profiles,panagoulia:2014.cluster.entropy.profile.compilation,sanders:2018.cluster.density.entropy.temperature.profiles.redshift.samples,kafer:2019.cluster.xray.sb.profiles.by.cc.status,ghirardini:2019.cluster.profiles.compilation.universal.fits,bartalucci:2023.xray.cluster.sb.profiles}, and see that this simple model produces a profile in all of these quantities which is very similar to SCC clusters.\footnote{Note that because this is a first study, and the best-observed clusters reside at $z\sim 0$, we model low-redshift systems here. In \papertwo, we expand upon the predictions for X-ray profiles at high redshifts $z\sim 1-2$, and show these are also similar to those observed for plausible assumptions. However we caution that any redshift extrapolation requires making (highly uncertain) assumptions for how CR injection rates and transport parameters, as well as CC gas properties, evolve.}

In other words, in this toy model, the CR-IC ``mimics'' a SCC in many key properties. But there is, by assumption, little actual thermal gas emission -- the ``apparent'' CC luminosity would be entirely dominated by the CR-IC emission. Using the standard definition of the cooling radius (where $t_{\rm cool} \lesssim 7\,$Gyr), we would infer an apparent CC radius $\sim 100\,$kpc, and apparent ``cooling luminosity'' within the CC radius of $L_{\rm X,\,cool}^{\rm apparent} \sim \dot{E}_{\rm cr,\,\ell} \sim 10^{44}\,{\rm erg\,s^{-1}}$.

\begin{figure*}
	\centering\includegraphics[width=1.0\textwidth]{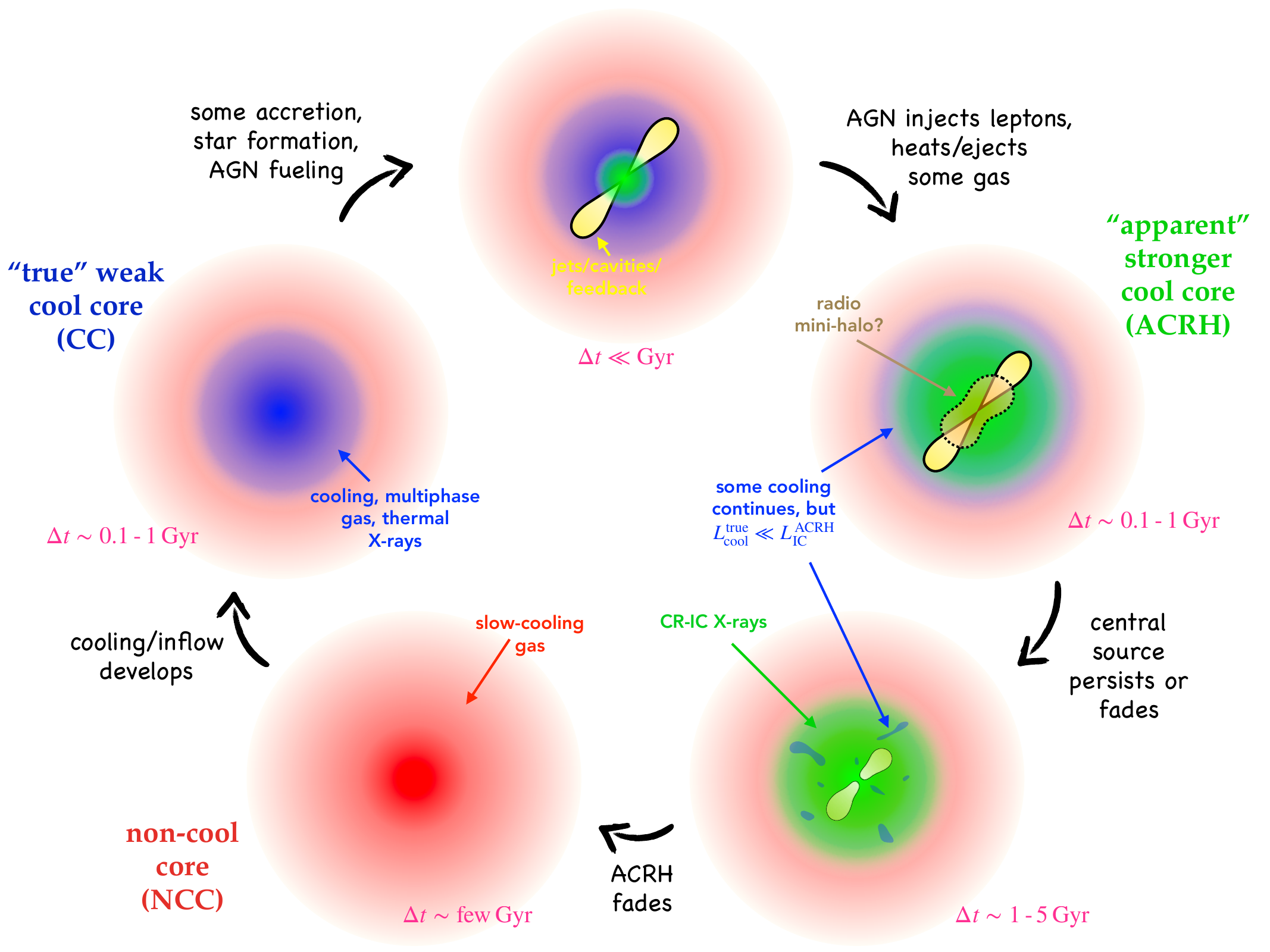} 
	\caption{Cartoon illustration of speculative time-series of events linking different cluster and radio source categories (\S~\ref{sec:cartoon}). Clockwise from bottom-left: An initial NCC develops some cooling and briefly resembles an ``uncontaminated'' CC, with inflows leading to some AGN activity which drives jets/cavities/feedback and injects CRs that form a growing central radio source and diffuse out to produce more extended emission similar to the CC profiles in Fig.~\ref{fig:profiles}. This boosts the apparent CC luminosity by the (potentially much larger) CR-IC X-ray luminosity, producing an ``apparent'' strong cool core (which is, in fact, an ancient cosmic-ray halo of low energy $\sim$ GeV leptons).  
	While the central AGN is strong it will contain an active radio galaxy, and potential radio mini-halo in the center surrounded by a larger radio-invisible (MHz) multi-phase CR-IC halo. Even after the central radio source weakens, the CR-IC halo can persist, slowly fading, for $\gtrsim$\,Gyr until the system again resembles a NCC. 
	We qualitatively estimate timescales for each phase but stress this is just a cartoon,  and the ``order of events'' could be more complicated than illustrated here.
	\label{fig:cartoon}}
\end{figure*}

\section{A New Interpretation of the Cooling Flow Problem}
\label{sec:cartoon}

The potential importance of IC emission in clusters with strong AGN suggests a rather radical re-interpretation of CC clusters and the cooling flow (CF) problem, which we illustrate heuristically in Fig.~\ref{fig:cartoon}.   X-ray line and multi-wavelength observations indicate there usually is \textit{some} multi-phase gas in CC clusters.  However, the amount of said gas, as well as the recent star formation rate and molecular and atomic and dust mass reservoirs of the CCs, are orders-of-magnitude smaller than the naively implied mass that should be present given the \textit{apparent} mass of CGM gas with cooling time less than the Hubble time (e.g.\ classical CF $\dot{M}_{\rm cool} \sim (\mu m_{p}/k_{B} T)\,L_{\rm X,\,cool}^{\rm apparent}(R < R_{\rm cool}) \sim 1000\,{\rm M_{\odot}\,yr^{-1}}\,(L_{\rm X,\,cool}^{\rm apparent} / 10^{44}\,{\rm erg\,s^{-1}})\,(T/{\rm keV})^{-1}$; see \citealt{fabian:1994.cluster.cooling.flows.review,fabian:2002.classical.cooling.flow.problem.obs.definition.missing.lum.profile.of.mdot.vs.r.wrong}).  SCCs are also ubiquitously associated with bright radio AGN, whose radio luminosity and apparent (X-ray cavity inferred) jet power appear to be tightly correlated with the apparent X-ray cooling luminosity $L_{\rm X,\,cool}$, implying feedback power and plausibly CR injection rates $\dot{E}_{\rm cr,\,\ell} \sim L_{\rm X,\,cool}^{\rm apparent}$ (see \citealt{birzan:2004.radio.cavity.jet.power.vs.xray.power.correlation,rafferty:2006.cluster.cavity.jet.power.vs.xray.luminosity.agn.feedback.arguments.but.accretion.rates.dont.match.obvious,rafferty:2007.xray.radio.cluster.data.compilation,nulsen:2007.cluster.jet.power.versus.cooling.luminosity.correlation.compilation,sun:2009.every.radio.agn.has.xray.cc.strong.lradio.lxray.connection.clusters.all.xray.agns.also.radio.agns,osullivan:2011.radio.jet.power.radio.emission.and.xray.cooling.luminosities,hlavacek.larrondo:2011.cluster.radio.cavity.jet.power.vs.cooling.xray.luminosity.extension,hlavacek.larrondi:2012.expanding.sample.luminous.clusters.xray.cooling.luminosity.vs.radio.cavity.jet.power}). 

In the ``traditional'' CF interpretation, this all arises via feedback power balancing radiative cooling (see \citealt{mcnamara:2007.agn.cooling.flow.review.cavity.jet.power.vs.xray.luminosity.scalings.emph.compilation,eckert:2021.agn.feedback.galaxy.groups.review.challenges.theory.obs.producing.realistic.coolcores} for reviews). The standard scenario involves a series of steps that are theoretically challenging to realize without fine tuning.   A CC begins to cool, producing for some reason a quasi-universal profile in $n_{\rm gas}$, $T$, $K$, $P$ (generally \textit{not} the same as the profiles predicted in hydrodynamic simulations; see \citealt{braspenning:2024.flamingo.simple.xray.modeling.sims.clusters.dont.reproduce.zdrops.other.cc.features,lehle:2024.simulation.cluster.profiles,nelson:2024.tng.cluster.sims.profiles.basic.properties,prunier:2024.tng.cluster.xray.cavities.properties}). Almost immediately, an AGN must be ``turned on'' by some tiny fraction of the CF cooling out of the halo
even though the vast majority of the cooling gas must not actually cool, to avoid over-predicting the cold gas reservoir. The AGN launches a jet, whose mechanical power (determined at horizon scales by physics of the disk geometry, BH spin, and magnetic field structure; e.g., \citealt{2011MNRAS.418L..79T}) somehow balances the ``global'' CF luminosity $L_{\rm X,\,cool}$ (almost all of which comes from the largest cooling radii $\gtrsim 100\,$kpc; \citealt{allen:2000.predicted.properties.xray.cooling.flows.not.there.missing.column.and.cold.gas,peterson:2003.cluster.cooling.flow.problem.missing.lowtemperature.gas.looks.multiphase.but.truncated.kev,hudson:2010.cool.core.cluster.review.central.properties.temperature.entropy.coolingtime.definition.coolcore.basic.scalings.size.luminosity.mdot}). This balance must be precise, to re-heat or mix the gas so as to suppress $\gtrsim 99\%$ of the naively implied cooling. This has led to serious theoretical challenges reproducing observations: the majority of simulations and models of CFs do not produce the large population of clusters that actually \textit{look like} SCCs -- rather AGN feedback, where effective, tends to over-heat or blow out gas from SCC centers, effectively quenching the cooling but giving rise to central density/entropy profiles that do not resemble the CC observations in Fig.~\ref{fig:profiles} (references above and \citealt{Martizzi2019, altamura:2023.large.volume.sims.independent.of.free.parameters.struggle.to.make.realistic.cc.clusters,su:2023.jet.quenching.criteria.vs.halo.mass.fire,su:2024.fire.jet.sim.using.acc.jet.prescriptions.from.cho.multiscale.experiments,gonzalez.villalba:2024.magneticum.cluster.cc.ncc.statistics.pred}).   It is, of course, very possible that this is just a limitation of existing simulations, but we suggest that it is also possible that the standard scenario requires revision.

\begin{figure}
	\centering\includegraphics[width=0.99\columnwidth]{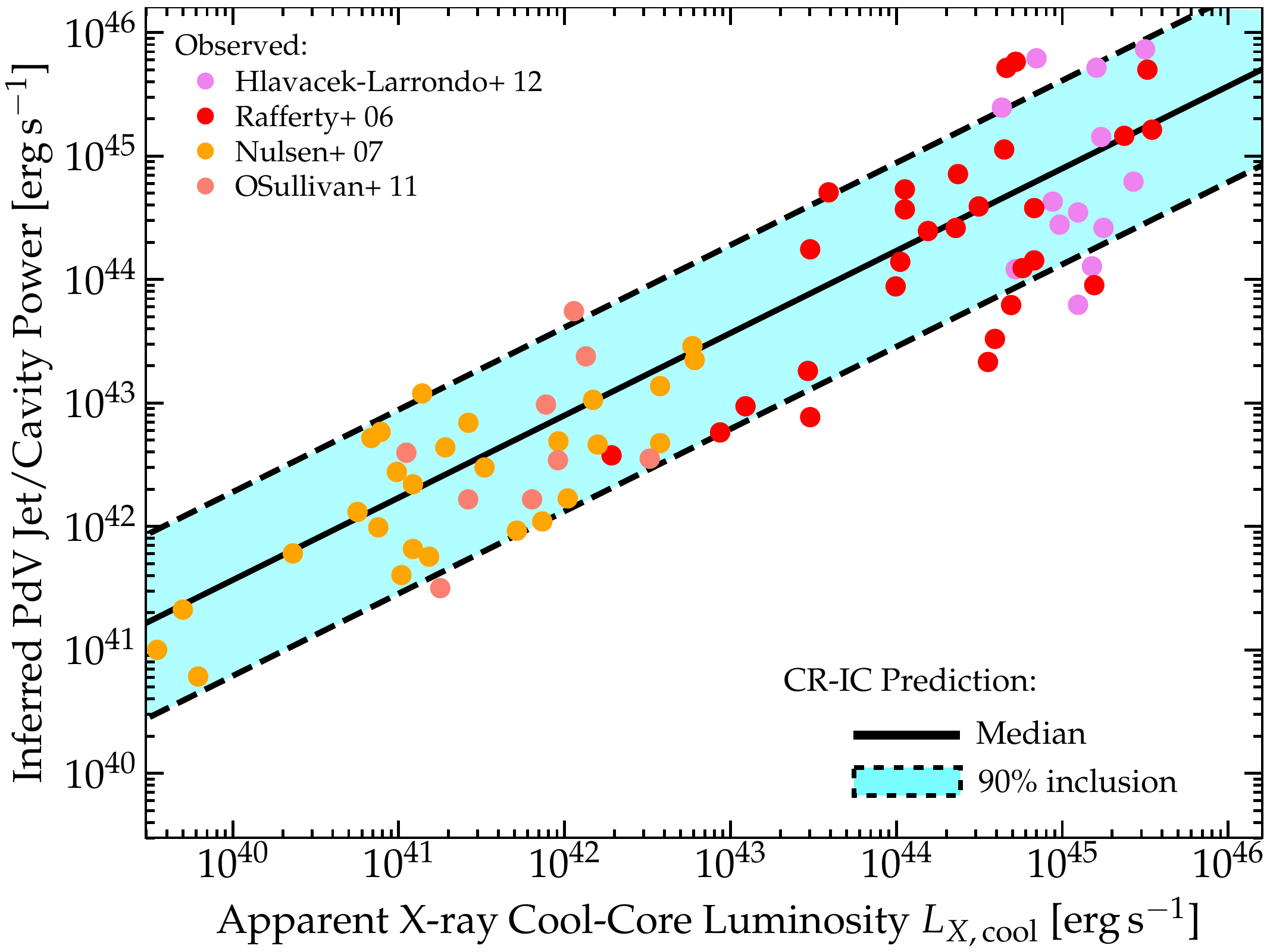} 
	\caption{Inferred ``cavity power'' $P_{\rm cav} \sim 4\,P_{\rm eff}\,V/t_{\rm buoy}$ (\S~\ref{sec:agn.XR}) versus X-ray cooling flow/cool-core luminosity $L_{\rm X,\,cool}$, which would be inferred for cavities in the CR-IC dominated models shown in Fig.~\ref{fig:profiles} (with different $\dot{E}_{\rm cr,\,\ell}$). 
	We show the median and $\sim 90\%$ spread (driven by the range of observed volume-filling factors of cavities). 
	We stress the predicted scaling (Eq.~\ref{eqn:pcav}) has no fitted/adjusted parameter here. 
	The correlation arises in the CR-IC model because both the X-ray inferred $P_{\rm cav}$ 
    and $L_{\rm X,\,cool}$ are functions of the \textit{same} leptonic CR energy density in the CC. 
	We compare to X-ray measurements in strong CC clusters (references shown). 
	In the CR-IC scenario this correlation and other radio-galaxy-CC correlations arise because we are effectively measuring two different functions of the CR energy density and plotting them against one another.  
	\label{fig:pjet.vs.lx}}
\end{figure}

Fig.~\ref{fig:cartoon} proposes an alternative. In a CC, some real cooling does occur, and this can help fuel AGN activity, just like in the traditional picture (this cooling is required by the existence of multiphase gas correlated with the CC properties; e.g., \citealt{Cavagnolo2008}). But once on, a luminous AGN produces a CR-IC halo whose properties are extremely similar to the observed properties of SCCs. The ``real'' thermal cooling luminosity is faint: the cooling luminosity that will be \textit{inferred} is $L_{\rm X,\,cool}^{\rm apparent} \sim \dot{E}_{\rm cr,\,\ell}$, powered primarily by CR-IC. 
This immediately explains a number of SCC properties more naturally than standard models, including:

\begin{itemize}
\item The ``classical CF'' problem is naturally resolved. $\dot{M}_{\rm cool} \propto L_{\rm X,\,cool}^{\rm apparent}$ is much larger than the ``real'' cooling luminosity (evident in colder multiphase gas) because the emission is boosted by CR-IC.   This also explains why CF models fit to SCC spectra require a ``cutoff'' at lower temperatures (there is not so much cooler gas):  the ``cutoff'' is set by the characteristic photon energy of the CR-IC emission relative to the virial temperature (Fig.~\ref{fig:mhalo}).  

\item The tight observed correlation between radio cavity/jet power, central AGN activity/radio luminosity, and $L_{\rm X,\,cool}^{\rm apparent}$ aquires a new, easier to understand and reproduce, interpretation: the same basic source (relativistic electrons from the AGN) is producing the apparent X-ray cooling luminosity and  the radio emission (Fig. \ref{fig:pjet.vs.lx}).

\item The characteristic CC radial profiles, in surface brightness, apparent density and temperature, entropy, and pressure (inferred from X-ray observations) all emerge naturally from CR-IC, as an immediate byproduct of how the CR-IC emissivity scales with radius. 
There are effectively just two important parameters, the lepton injection rate $\dot{E}_{\rm cr,\,\ell}$ and CR streaming/diffusion speed $v_{\rm st,eff}$, which control the normalization, i.e.\ apparent strength of the CC. The characteristic radii $\sim 100\,$kpc of CCs (and their scaling with $L_{\rm X,\,cool}^{\rm apparent}$) also emerge as a result of the CR-IC loss timescale.
\end{itemize}

The re-interpretation of CC clusters suggested in Figure \ref{fig:cartoon} removes the fine-tuning challenges from the traditional interpretation. For an almost arbitrary AGN luminosity above a certain threshold (set by the true thermal emission), the AGN-apparent X-ray cooling correlations and apparent CC profiles will emerge automatically from that AGN itself.

\subsection{The X-ray AGN Connection}
\label{sec:agn.XR}

For illustration, consider a couple examples of the AGN-X-ray correlations  interpreted as evidence for feedback regulation of hot halos. There is a well-known correlation between cooling luminosity $L_{\rm X,\,cool}^{\rm apparent}$ and central AGN activity, radio luminosity, or point-source AGN X-ray luminosity \citep{rafferty:2007.xray.radio.cluster.data.compilation,mcnamara:2007.agn.cooling.flow.review.cavity.jet.power.vs.xray.luminosity.scalings.emph.compilation,sun:2009.every.radio.agn.has.xray.cc.strong.lradio.lxray.connection.clusters.all.xray.agns.also.radio.agns,osullivan:2011.radio.jet.power.radio.emission.and.xray.cooling.luminosities,hlavacek.larrondi:2012.expanding.sample.luminous.clusters.xray.cooling.luminosity.vs.radio.cavity.jet.power,liu:2024.strong.lradio.lxray.connection.luminous.cool.cores.strong.central.xray.only.when.strong.radio.source}. If CR-IC boosts the SCC, then this correlation is readily explained: $L_{\rm X,\,cool}^{\rm apparent} \sim \dot{E}_{\rm cr,\,\ell}$ is itself injected by the AGN, and the AGN radio emission traces the same leptonic source, just with radio tracing higher-energy CRs injected more recently, and the X-rays older CRs injected $\sim$\,Gyr prior (this difference in timescale implies that there should be scatter in the observed correlation related to the current jet power [relevant for radio] vs the Gyr-averaged jet power [relevant for IC X-rays]).

Fig.~\ref{fig:pjet.vs.lx} shows the correlation between AGN apparent ``cavity'' or ``jet'' power $P_{\rm cav}$ and $L_{\rm X,\,cool}^{\rm apparent}$, often cited as the most important evidence for the interpretation of the CF problem as a fine-tuned balance between cooling and AGN heating \citep{rafferty:2006.cluster.cavity.jet.power.vs.xray.luminosity.agn.feedback.arguments.but.accretion.rates.dont.match.obvious,nulsen:2007.cluster.jet.power.versus.cooling.luminosity.correlation.compilation,osullivan:2011.radio.jet.power.radio.emission.and.xray.cooling.luminosities,hlavacek.larrondi:2012.expanding.sample.luminous.clusters.xray.cooling.luminosity.vs.radio.cavity.jet.power}. Specifically, from the models in Fig.~\ref{fig:pjet.vs.lx}, we can calculate $L_{\rm X,\,cool}$ as defined observationally ($L_{X}$ where $t_{\rm cool}<7\,$Gyr interpreting the emission as thermal emission). $P_{\rm cav}$ is observationally defined by the apparent ``PdV work'' associated with X-ray cavities, specifically $P_{\rm cav} \equiv 4\,p_{\rm ext}\,V_{\rm cav}/t_{\rm buoy}$ where $p_{\rm ext} = P_{X}$ is the apparent (X-ray fit/inferred) hot gas pressure outside the cavity ($P_{X}$ in Fig.~\ref{fig:profiles} at the cavity distance $R$), and $t_{\rm buoy} \equiv R/v_{\rm buoy} \equiv R \sqrt{S\,C/2\,g\,V_{\rm cav}}$ an estimate of the buoyancy time (with $S\equiv \pi R_{\rm cav}^{2}$, $V_{\rm cav} \equiv (4\pi/3)\,R_{\rm cav}^{3}$, $C\equiv 0.75$, $g\equiv 2\,\sigma^{2}_{\ast,\,\rm gal}/R$; $\sigma^{2}_{\ast,\,\rm gal}\equiv 280\,{\rm km\,s^{-1}}$ defined in \citealt{birzan:2004.radio.cavity.jet.power.vs.xray.power.correlation}). 

If we define the dimensionless $\xi_{\rm vol} \equiv (2\,R_{\rm cav}/R)^{7/2}/(R_{\rm cool}/2\,R)$ (scaling roughly with the volume-filling factor of the cavities), and take the values of $R_{\rm cav}$ and $R$ quoted in the papers above (where the values were measured), we obtain the prediction for $P_{\rm cav}$ versus $L_{\rm X,\,cool}$ shown in Fig.~\ref{fig:pjet.vs.lx}, in terms of quantities predicted in Fig.~\ref{fig:profiles}.  A detailed analytic derivation is given in \papertwo, but the predicted median scaling is: 
\begin{align}
\label{eqn:pcav} \frac{P_{\rm cav}}{10^{44}\,{\rm erg\,s^{-1}}} \sim \frac{1.5\, \xi_{\rm vol}}{(1+z)}\,\left( \frac{L_{\rm X,\,cool}}{10^{44}\,{\rm erg\,s^{-1}}} \right)^{2/3} \ .
\end{align} 
where our definition of $\xi_{\rm vol}$ is such that the median (where measured) is $\sim 1$ with a $\approx 0.4\,$dex $1\sigma$ spread.  

We stress that this is a \textit{prediction}, with no free parameter representing some unknown efficiency or conversion of accretion energy to CRs or heat or X-rays. The clean prediction arises because (1) $L_{\rm X,\,cool}$, if determined by CR-IC, is simply set by the CR energy density (since that determines both the CR-IC emissivity and spectrum, hence apparent cooling time) and size of the region, while (2) $P_{\rm cav} \propto P_{X}\,V_{\rm cav}/t_{\rm buoy}$ just depends on the apparent X-ray pressure (which is \textit{also} just a function of the CR energy density for the same reasons) and size of the region. In the CR-IC interpretation, we are just measuring two different functions of the CR energy density and plotting them against one another.

\begin{figure}
	\centering\includegraphics[width=0.99\columnwidth]{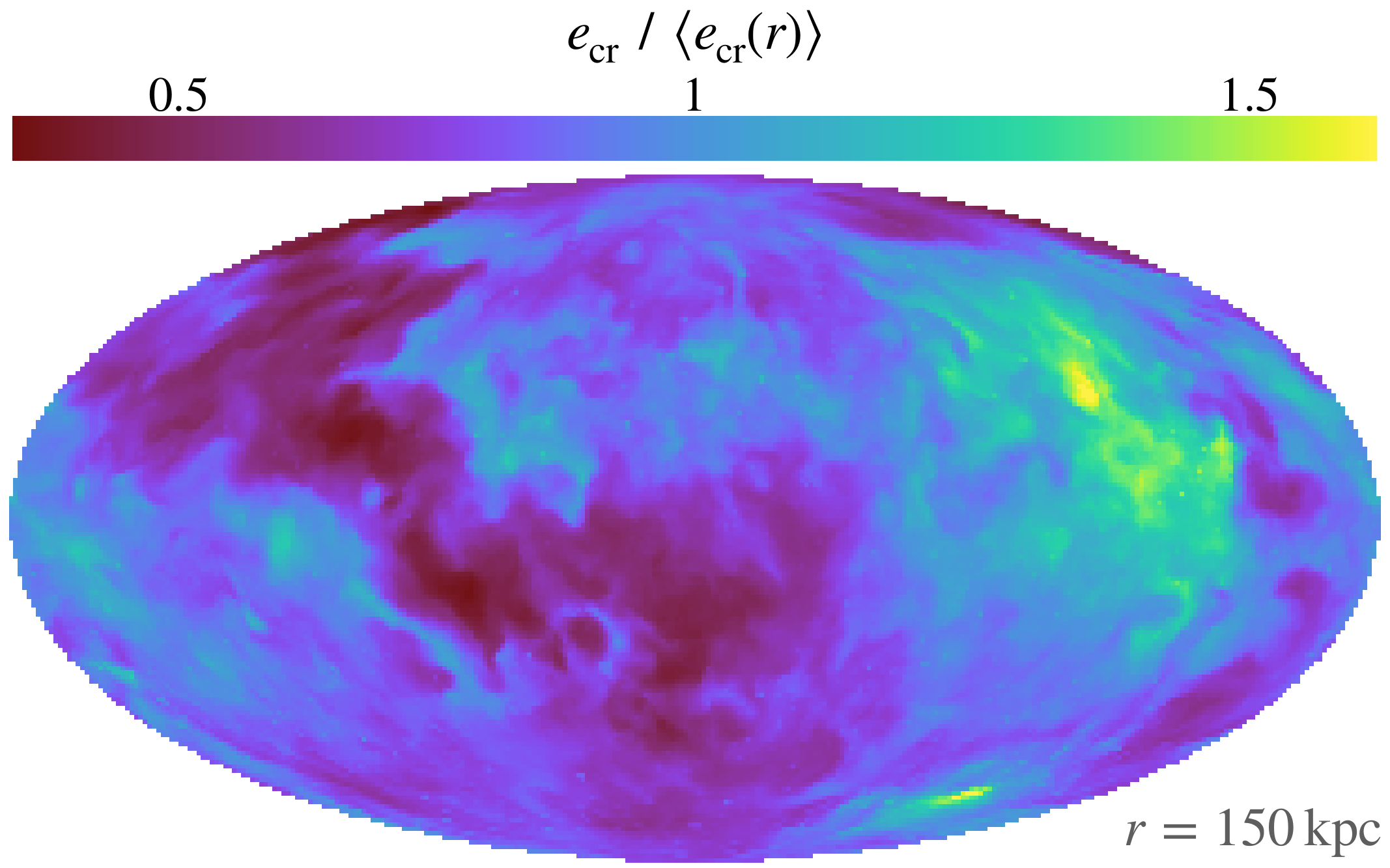} 
	\caption{Illustration of CR energy density fluctuations, which would seed corresponding surface-brightness fluctuations, in a CR-IC scenario. 
	{\em Top:} Example Mollweide projection of the $\sim$\,GeV CR energy density (proportional to CR-IC X-ray emissivity) in a thin shell at $r=150\,$kpc, in a simulation of a Milky Way-mass halo with similar diffusion/streaming speeds to the analytic model in Fig.~\ref{fig:profiles} (adapted from \citealt{ji:fire.cr.cgm}). With $\gtrsim$\,Gyr transport times at these radii, CRs are not distributed perfectly smoothly and can produce $\sim10\%$ brightness fluctuations similar to those observed (in addition to there being intrinsic gas density fluctuations).
	\label{fig:pressure.fluct}}
\end{figure}

\subsection{X-ray Substructure and Multi-phase Gas}
\label{sec:fluct}

Although the idealized analytic model considered here assumes spherical symmetry, with steady-state CR injection, in reality there is likely to be X-ray sub-structure in CR-IC halos. Fig.~\ref{fig:pressure.fluct} illustrates  this in a related context, showing the predicted CR energy density in a shell at $\sim 150\,$kpc from the central galaxy in the CGM in a cosmological CR-MHD galaxy formation simulation (but in this case in a Milky Way-mass halo $\sim 10^{12}\,M_{\odot}$, since that is where we have simulations with the appropriate physics). 
This simulation (at $z=0$, shown) has a strongly CR pressure-dominated halo and the predicted CR-IC X-ray luminosity (proportional to the CR energy density) dominates over thermal X-ray emission at this radius \citep{hopkins:2025.crs.inverse.compton.cgm.explain.erosita.soft.xray.halos,lu:2025.cr.transport.models.vs.uv.xray.obs.w.cric}. 
The details will differ in cluster cores, but the point is that fluctuations of tens of percent and up to order-unity decrements in the implied X-ray emission can appear in fluctuations with sizes $\mathcal{O}(r)$. These can arise owing to differential losses (e.g.\ different ``PdV work'' done along different trajectories) and global compression/turbulence (including phenomena known already to exist in CCs, like cold fronts, sloshing, shocks, etc.), but they must be present in large part simply due to transport physics: recall at the radii of interest, the CR transport times are by definition $\sim$\,Gyr, comparable to or even longer than global dynamical/free-fall times in CCs. So there is no reason for them to be perfectly smooth (finite travel-time and variable injection alone can introduce potentially interesting substructure; see Ponnada et al., in prep.). Moreover, a number of authors have pointed out that CR streaming in outflows can produce coherent overdensity structures (e.g.\ arcs or ``harps''; \citealt{huang.davis:2021.cr.staircase.in.outflows,tsung:2021.cr.outflows.staircase,quataert:2021.cr.outflows.diffusion.staircase,ruszkowski.pfrommer:cr.review.broad.cr.physics}). These are all similar to the sorts of structures seen in X-ray surface brightness fluctuations in CCs \citep[e.g.][]{zhuravleva:2018.cluster.turb.props.from.xrays}. 
And in most of the diffuse volume of CCs, low-frequency radio and X-ray surface brightness fluctuations are clearly observed to be positively correlated \citep{giacintucci:2011.cluster.minihalo.fills.cool.core.morphologically.similar,giancintucci:2019.expanding.radio.cluster.minihalo.sample.no.good.corr.total.cluster.mass.or.total.cluster.xray.but.very.strong.corr.cooling.radius.xray.luminosity.consistent.with.linear.standard.correlation,bravi:2016.minihalo.luminosity.strong.corr.xray.luminosity.clusters,balboni:2024.lofar.xray.surface.brightess.corr.indiv.halos.well.corr.positive.as.expected.for.ic.same.particles.but.sublinear.because.spread.by.B.scatter,riseley:2023.cluster.radio.minihalo.correlated.xray.brightness.disturbed.re.energized.example,riseley:2024.large.minihalo.reenergized.with.merger.but.center.shows.strong.radio.xray.with.steeper.radio.hotter.gas.and.local.ir.ix.corr.as.expected.for.ic.nearly.linear,vanweeren:2024.perseus.giant.radio.halo.filled.electrons.just.tiny.fraction.high.energy} -- the exception being central X-ray cavities which are part of the ``injection zone'' (as we state in \S~\ref{sec:basic}, these will be X-ray dim, in both CR-IC and thermal emission, in these models, relative to their radio, but observationally are restricted to the young/active jet regions). 
The possibility of fluctuations in the CR-IC emission cautions against always interpreting observed SB fluctuations one-to-one as thermal emission fluctuations: the true amplitude of the thermal emission fluctuations may be different in a CR-IC-boosted CC. 

Observations show that CC clusters contain significantly more multiphase gas than NCC clusters \citep{Cavagnolo2008}, although far less than one might anticipate given the cooling rate of the hot gas.   This implies that the (intentionally) most extreme version of the CR-IC model shown in Figure \ref{fig:profiles} -- in which the thermal emission is that of a NCC cluster and the remaining emission is CR-IC -- is probably not viable.   Instead, our proposal is more analogous to that shown in Figure \ref{fig:cartoon}, in which observed CC clusters have somewhat denser gas than NCC systems, which is precisely what triggers the luminous AGN and its associated CR-IC X-ray emission. 
 The ``bath'' of CRs interacting with this multiphase gas may in fact naturally explain a number of otherwise puzzling aspects of the ionization and line excitation of observed atomic and molecular gas in CCs, as emphasized in earlier work \citep{ferland:2009.particle.ionization.needed.for.molecular.line.emission.in.cc.perseus,mittal:2012.perseus.core.atomic.gas.dust.mass.cold.atomic.lines.to.25.kpc.densities.temperatures.ionization,canning:2016.Halpha.CII.OI.CO.modeling.excitation.line.ratios.skin.depths.extreme.excitation.cold.atomic.gas}; we will study this in detail in the future.   Within the CR-dominated central CC, we still predict multi-phase gas should exist, and at least some of the decrease in $T$ at small radii is likely real. Indeed, CRs actually promote the formation, stability and longevity of multi-phase cooler gas \citep{ji:fire.cr.cgm,hopkins:2020.cr.outflows.to.mpc.scales,ji:20.virial.shocks.suppressed.cr.dominated.halos,hopkins:2020.cr.transport.model.fx.galform,butsky:2020.cr.fx.thermal.instab.cgm,weber:2025.cr.thermal.instab.cgm.fx.dept.transport.like.butsky.study}, governed globally by buoyant dynamics \citep{hopkins:2020.cr.outflows.to.mpc.scales}.

\begin{figure}
	\centering\includegraphics[width=1.02\columnwidth]{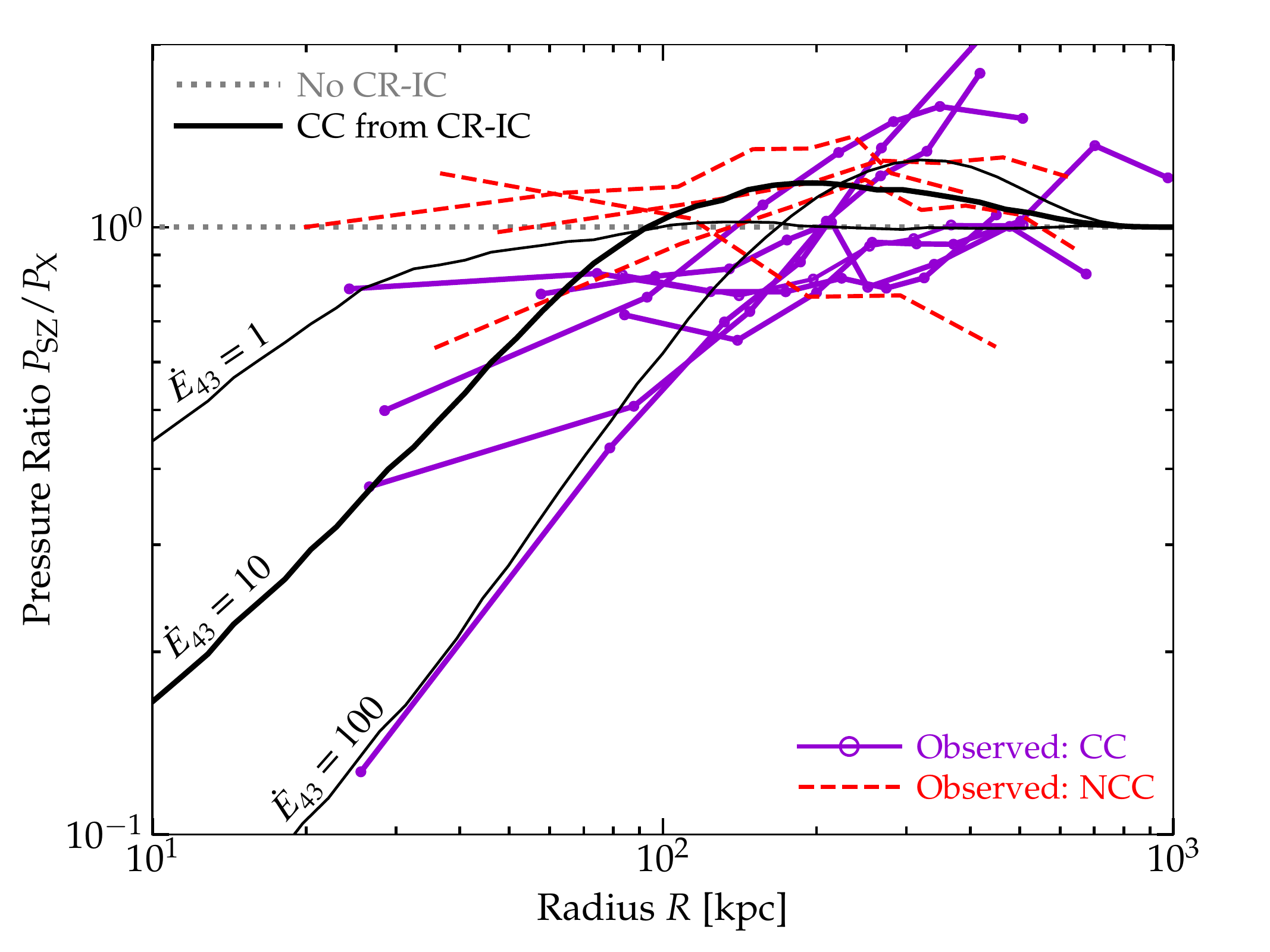} 
	\caption{Calculated ratio of the true gas pressure $P_{\rm true} = n_{\rm true}\,k_{B} T_{\rm true}$ to the \textit{apparent} (X-ray inferred) gas pressure $P_{\rm X} \approx P_{\rm eff} = n_{\rm eff}\,k_{B}\,T_{\rm eff}$ for the same toy-model profiles shown in Fig.~\ref{fig:profiles}. 
	Since SZ should trace the true pressure $P_{\rm SZ} \approx P_{\rm true}$, we plot the observed ratio $P_{\rm SZ}/P_{\rm X} \approx P_{\rm true} / P_{\rm eff}$, for CC clusters and NCC clusters with high-resolution SZ observations probing the central regions (\S~\ref{sec:sz}; \citealt{romero:2017.cluster.pressure.profiles.highres.sz.xray.cool.cores.show.central.pressure.deficit}). At large radii there is no effect of CRs; at smaller radii the CR-IC emission leads to excess X-ray surface brightness which causes an over-estimate of $n$ and hence $P_{\rm X} > P_{\rm SZ}$. Remarkably, this toy-model example appears very similar to the median behavior seen in CC clusters, where most of the observed CC systems tentatively exhibit an SZ ``pressure deficit'' in their central regions, while almost no NCC clusters exhibit such a deficit. 
	\label{fig:sz}}
\end{figure}

\section{Observable Tests: The SZ Effect and Cool-Core Center Metallicities}
\label{sec:obs}

While the scenario in \S~\ref{sec:cartoon} can explain many observations and theoretical challenges, it does not, by itself, prove that the emission in the centers of CC clusters has significant contributions from CR-IC. Testing this in observations is surprisingly difficult. As shown explicitly in Figure \ref{fig:spectrum}, at the radii of interest, the continuum X-ray emission appears thermal.  In addition, the co-spatial radio, hard X-ray IC, and $\gamma$-ray emission produced by the aged GeV CRs are generally undetectable in any current observations or planned future missions over the next several decades (more details in \papertwo). 

In principle, comparing X-ray inferred hydrostatic mass profiles $M(r)$ (or potentials $\phi$) at small radii in CC clusters to profiles measured by stellar dynamics or gravitational lensing could indirectly constrain CR-IC emission in X-rays. The X-ray inferred mass profile could be biased owing to (1) the CR-IC emission giving a different inferred X-ray temperature and density, and (2) the CR lepton pressure itself being dynamically significant at small radii. For example, Fig~\ref{fig:profiles} shows that in the intentionally extreme case of CR-IC mimicking a SCC in a NCC cluster, the CR pressure is larger than the true gas pressure interior to $\lesssim 10$\,kpc (this is not true at larger radii).  In \papertwo\ we show that for all but the most extreme CCs, the effect of the CR-IC on $\phi$ or $M(r)$ is modest, 
owing to (1) the fact that potential reconstruction is most sensitive to temperature, which is more weakly influenced by CR-IC, and (2) the different effects of CR-IC (on inferred density profile slopes vs.\ absolute temperature) have opposite signs and offset one another.  Initial estimates of the deviation between X-ray and stellar-dynamical measurements of $\phi$ in the best-constrained  weak CCs (M87+Fornax) were $\sim 10-50\%$ \citep{2008MNRAS.388.1062C}. However, better modeling and new IFU datasets have changed the M87 stellar dynamical modeling significantly such that the two methods now disagree by a factor up to $\sim 2$ in $\phi$ at $\lesssim 10$\,kpc \citep{gebhardt:2009.bh.mass.revision.new.models.m87,murphy:2011.new.m87.models,liepold:2023.m87.dynamical.masses}, and in SCCs the deviations can be much larger \citep{zhang:2008.cluster.profiles.lensing.xray.good.down.to.0pt2.r500.inner.makes.scatter.much.larger.biases.cosmological.measurements,newman:2013.cluster.mass.profiles.multi.method,simet:2017.weak.lensing.xray.cosmology.masses.agreement.but.large.radii,pratt:2022.cluster.density.profiles.also.need.to.excise.cores.to.get.clean.lx.t.mass.relations,allingham:2023.clusters.kinematic.lensing.vs.xray.mass.profiles.large.disagreement.qualitative.similar.nfw.profiles}. 
It is also important to stress that because the effects of CR-IC are confined to cluster cores, our predictions have no effect on larger-scale cluster properties and comparisons of X-ray to strong and/or weak lensing-inferred masses at large radii in clusters (such as, e.g., \citealt{Newman2013}).   For the same reason, our predictions specifically do not affect the use of clusters for cosmological applications.

There are, however, two effects which could directly detect the importance of CR-IC emission, which we now discuss.

\subsection{Sunyaev-Zeldovich}
\label{sec:sz}

The Sunyaev-Zeldovich (SZ)  effect gives a clean probe of the true thermal gas pressure in the cluster center $P_{\rm SZ}$ (the traditional measurement being sensitive to weak scattering by the much more numerous thermal electrons, scaling as $n_{e}\,T_{e}$, as compared to the rare CR-IC strong-scattering). In the models here, independent of how strong CR-IC emission is, one should still have $P_{\rm SZ} \approx P_{\rm thermal,\,true} \approx n_{e}\,k_{B}\,T$. 
However, as shown in Fig.~\ref{fig:profiles}, if CR-IC adds significantly to the central X-ray surface brightness, then it will artificially boost the X-ray inferred central pressure, $P_{\rm X} \approx P_{\rm thermal}^{\rm apparent,\,XR} \sim n_{\rm apparent}\,k_{B}\,T_{\rm apparent}  > P_{\rm thermal,\,true}$. 
So if CR-IC is negligible, $P_{\rm SZ} \approx P_{\rm X}$, while if CR-IC dominates the X-ray emission, $P_{\rm SZ} < P_{\rm X}$. We stress this is independent of any physics that causes changes in the true pressure, like CR pressure support, deficits in bubbles/cavities, etc., as well as many details of the observations, so long as they are binned to similar scales (and not e.g.\ extrapolated using some assumed universal/similar functional forms). 
Fig.~\ref{fig:sz} shows this prediction for the same specific model in Figure \ref{fig:profiles}. 
Noting that the primary effect of CR-IC is to boost the apparent gas density (since surface brightness $\propto n_{\rm gas}^{2}$, while it is a weaker function of temperature), the expected decrement $P_{\rm SZ}/P_{\rm X}$ should scale roughly as $\sim (I_{\rm thermal}/I_{\rm total})^{1/2}$.

Observationally, it is challenging to measure the radii where the predicted $P_{\rm SZ} \ll P_{\rm X}$, because they are confined to cluster cores (and one needs to avoid contamination by the central AGN as well), requiring few arcsecond angular resolution (much finer than most cluster SZ studies using Bolocam/ACT/SPT/Planck). However MUSTANG (and perhaps NIKA2) does reach sufficient resolution, and there is one study in \citet{romero:2017.cluster.pressure.profiles.highres.sz.xray.cool.cores.show.central.pressure.deficit} of $P_{\rm SZ}$ and $P_{\rm X}$ on these scales with a small sample of $14$ clusters, of which $7$ are relaxed CC clusters. We show those in Fig.~\ref{fig:sz}. Remarkably, as the authors concluded,  almost every CC in their sample is consistent with some suppression $P_{\rm SZ} \ll P_{\rm X}$ in the CC centers at $R \ll 100\,$kpc, consistent with our models if we take $\dot{E}_{43}\sim 1-10$ to be in the same range as $L_{\rm X,\,cool}^{\rm apparent}$ observed in the sample of CCs they study. \citet{romero:2017.cluster.pressure.profiles.highres.sz.xray.cool.cores.show.central.pressure.deficit} fit a median $P_{\rm SZ}/P_{\rm X} \sim 1/3$ in the central $\lesssim 30\,$kpc in SCCs, consistent with $\sim 90\%$ of the emission from CR-IC (though with a broad range $\sim 40-98\%$ implied by the SCC sample). 
Meanwhile the NCC systems in \citet{romero:2017.cluster.pressure.profiles.highres.sz.xray.cool.cores.show.central.pressure.deficit} show no evidence for central CR-IC -- i.e.\ have $P_{\rm X} \approx P_{\rm SZ}$ in their centers (consistent with most of the emission being thermal).

\begin{figure}
	\centering\includegraphics[width=0.49\textwidth]{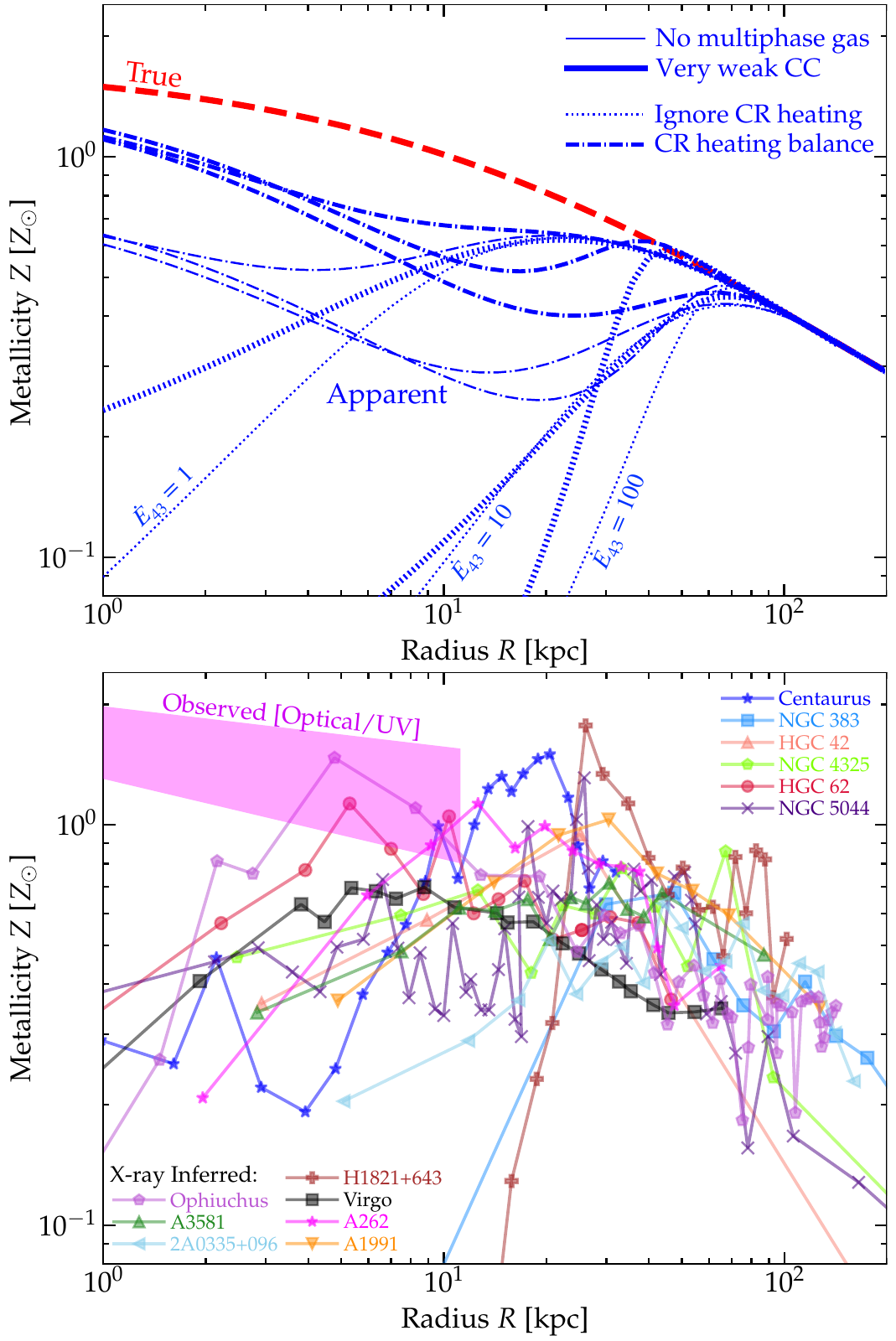} 
	\caption{{\em Top:} Toy-model illustration of how CR-IC can influence spectral estimates of X-ray observed metallicities ($Z$) in cluster centers ($R \lesssim 30\,$kpc; \S~\ref{sec:metal}). 
	We take the cluster models from Fig.~\ref{fig:spectrum}-\ref{fig:profiles}, assume a true $Z$ anchored to optical/UV measurements at $\lesssim 10\,$kpc and X-ray estimates at $>100\,$kpc, and estimate the ``apparent'' $Z$ from the iron line complex in the model spectra (assuming perfect measurements, deconvolution, and temperature recovery) assuming CR-IC with the given $\dot{E}_{43} \sim 1-100$ and some thermal emission. 
	For the latter we contrast a model with \textit{no} multi-phase gas (a NCC profile as Fig.~\ref{fig:profiles}; \textit{thin lines}), and one with a very weak ``true'' CC (WCC profile from \citet{mcdonald:2013.cluster.gas.profiles} with ``true'' CC cooling luminosity $\sim 10^{42}\,{\rm erg\,s^{-1}}$). We also contrast models ignoring any CR heating/ionization/gas interactions (pure-IC; {\em dotted}) or assuming all CR-gas heating is re-radiated at the virial temperature (``heating balance''; {\em dot-dashed}). 
	The predictions are sensitive to these and other detailed assumptions (e.g.\ ``drops'' only appear at $\lesssim 10\,$kpc in the most extreme models neglecting CR heating and stimulated emission), but in all cases CR-IC boosts the continuum and dilutes the lines relative to pure thermal emission, causing some under-estimate of $Z$ in the central $\ll 10-50\,$kpc. Thus the X-ray inferred $Z$ rises more slowly than the true value. 
	{\em Bottom:} Observed X-ray $Z$ profiles of CC groups/clusters with spatial resolution reaching $\lesssim 5\,$kpc (\textit{lines}, labeled). 
	We compare (\textit{pink shaded}) the observed gas (and stellar) phase metallicities of BCGs (including the same clusters) from ISM measurements with standard optical/UV estimators. 
	High-spatial-resolution observations of CCs appear consistent with an apparent metallicity suppression in the X-ray relative to optical/UV. 
	\label{fig:Z}}
\end{figure}

\subsection{Metal Line Properties}
\label{sec:metal}

A second test for CR-IC emission is in detailed metal-line measurements in CC clusters.  
CR-IC from upscattering the CMB  has a strikingly thermal continuum shape, but no lines, whereas thermal cooling in X-rays is often dominated (especially at lower temperatures $T \lesssim 2-3\,$keV) by strong $\sim$\,keV line emission.  If CR-IC is important for the continuum emission, the zeroth order effect is that the inferred metallicities from X-ray spectroscopy will be biased lower. In more detail, if CR-IC contributes to the X-ray SB in SCC centers, what would be observed in high spatial (arcsecond) resolution spectroscopy would be a mix of emission from (1) CR-IC from the CMB (thermal continuum-like but without strong lines) + (2) hot gas cooling (with lines) + (3) multi-phase cooler gas emission (clearly still present in CCs, per \S~\ref{sec:cartoon}, but in relatively small amounts), with lines + (4) excited line emission from CR interactions with the metals in those gas phases. 
This is quite complicated to model in general, especially not knowing either the true properties of the cooler X-ray emitting gas, nor the details of CR line excitation at X-ray wavelengths. The CR ionization cross-sections have large uncertainties for the partially-ionized metals of interest, and the stimulated excitation from secondary and Coulomb processes are even more poorly understood. 

However, we can draw some qualitative conclusions. If the SB at $\sim 1\,$keV has a significant contribution from CR-IC, then even with some amount of additional line excitation from CRs, the net effect will tend to be adding continuum to the spectrum, on top of the line-emission+continuum spectrum from multi-phase gas (modified by CRs). This will tend to dilute the line equivalent widths. If the X-ray spectra are fit to \textit{single-temperature, single-phase, single-metallicity} models around that wavelength range, as is common, they can still be well-fit with a thermal spectrum at some effective temperature (like in Fig.~\ref{fig:profiles}), but with a lower inferred metallicity $Z_{\rm obs}^{\rm 1T\,fit}$ compared to the true value $Z_{\rm true}$. It will also influence line ratios and linewidths (e.g.\ influencing the difference between spectral-shape and line-ratio estimates of X-ray temperature), but we show in detailed modeling in \papertwo\ that (in part because of the temperature sensitivity of the line-ratios) deviations even in single-temperature best-fits are very small, and are strictly degenerate with standard multi-temperature fits to the X-ray spectra (with an overall diluted metallicity). 

This is illustrated qualitatively in Fig.~\ref{fig:Z}. We assume a $Z_{\rm true}$ profile anchored to independent optical/UV measurements of BCG gas-phase metallicities and their gradients at $\le 10\,$kpc \citep{ellison:2009.mzr.clusters.cluster.bcgs.gas.phase.supersolar.even.moreso.if.bcg.or.cluster.center,manucci:2010.fundamental.sf.metallicity.relation.strong.sfr.dependence.low.mass.weak.high.mass,maier:2022.bcg.gas.phase.ism.abundances.supersolar.gradients.consistent.others,castignani:2022.bcg.gasphase.sfrs.metallicities.most.sf.bcgs.still.ism.supersolar,zinchenko:2024.manga.stellar.and.gas.metallicities.same.galaxies.differences.overall.very.consistent} and cluster X-ray compilations at $>100\,$kpc \citep{mernier:2017.cluster.metallicity.profiles.mean,lovisari.reiprich:2019.cluster.metallicity.profiles.compilation.larger.radii}, and qualitatively estimate the ``apparent'' inferred profile from Fe line strengths\footnote{Rather than model any one specific observation, we assume a simple proxy for illustration: $Z_{\rm obs}^{\rm fit} \approx Z_{\rm true}\,({\rm EW}_{\rm obs}/{\rm EW}_{\rm therm}$), where ${\rm EW}_{\rm obs/therm}$ is the equivalent width of the Fe line complex summed between $6-7$\,keV (assuming perfect resolution/signal to noise/background/continuum fitting/deprojection/temperature recovery/etc., and ignoring the temperature-metallicity degeneracy) from the total spectrum (``obs'') or the spectrum of just the true thermal emission (``true'').} assuming the same cluster model as Figs.~\ref{fig:spectrum}-\ref{fig:sz}, but varying: (1) $\dot{E}_{43}\sim1-100$; (2) whether we include \textit{no} multi-phase gas (only the NCC gas profile from Fig.~\ref{fig:profiles}) or assume there is a very weak ``true'' CC (taking $n_{\rm gas}$, $T_{\rm gas}$ from WCC fits); and (3) whether we ignore CR heating/excitation/ionization or mock some CR heating balance (crudely approximated by assuming the thermalized CR heating terms, \S~\ref{sec:model}, are re-radiated by gas at the virial temperature). We stress this is entirely qualitative and not a rigorous prediction in any sense (the real multi-phase structure would be very complex, especially in the CR-heated and CR-pressure-dominated regime, per \citealt{butsky:2020.cr.fx.thermal.instab.cgm,weber:2025.cr.thermal.instab.cgm.fx.dept.transport.like.butsky.study}), but we see that the profiles all show some $Z$ suppression (though how much varies widely) interior to $\ll 10-50\,$kpc, with the few most extreme model cases we consider even showing central $Z$ ``drops.'' We compare a number of CC profiles for which single-temperature fit, de-projected X-ray spectral metallicities are available at $\le 5\,$kpc spatial resolution \citep{matsushita:2002.m87.virgo.cluster.obs.metallicity.drop.temperature.fitting.challenges,rasmussen.ponman:2007.metallicity.temperature.profiles.groups,komiyama:2009.5044.metallicity.profile,million:2010.ophiuchus.central.metallicity.profile,murakami:2011.fornax.metallicity.profile.suzaku.xmm,walker:2013.centaurus.cluster.few.other.large.radii.entropy.metallicity.profiles,panagoulia:2015.cluster.metal.profiles.with.drops,sanders:2016.centaurus.metallicity.profiles.zdrop,gatuzz:2023.centaurus.metal.profile.outer,gatuzz:2023.ophiuchus.metallicity.profile,russell:2024.highres.qso.xray.coolingflow.zdrop.strong.entropydrop.verysmall.but.superluminous.coolingflow.much.larger.than.can.be.explained.physical.cooling.mechanisms}. 
As noted in those papers, even extreme ``drops'' are often seen in SCCs at sufficiently high ($<10\,$kpc, or $\sim1^{\prime\prime}$) spatial/angular resolution, though some of these remain controversial.
Much more importantly for our purposes, almost all cases (even those without any evidence for drops) are consistent with central \textit{suppression} ($Z \lesssim 1-2\,Z_{\odot}$ at $R < 10\,$kpc, with a relatively slow rise or ``plateau'' in $Z$ at these radii), in contradiction to the UV/optical metallicity measurements of gas \textit{at the same radii} (the $\sim 90\%$ inclusion interval of BCG gas-phase metallicity profiles from optical/UV measurements compiled above is shown, for comparison), which universally show $Z > Z_{\odot}$ and a steeper rise in $Z$ down to $<10\,$kpc.

We again stress that modeling the X-ray  spectra to constrain metallicities is highly degenerate, especially in low-resolution spectra from instruments like Chandra and XMM \citep{avestruz:2014.cluster.mocks.from.sims.sensitivity.temperature.measurements}. 
Indeed, it is well-known that even basic results from fitting line ratios in SCC centers, like the central metallicity or line-ratio inferred temperature, can differ systematically by up to an order-of-magnitude even for the same SCC based on fits to (a) different data from different instruments with different sensitivities and spectral resolutions \citep{ghizzardi:2021.iron.cluster.profiles.sensitivity.of.fitting.different.lines.different.metallicities,zhuhone:2023.cluster.temperature.fitting.sensitivities.simulations}; (b) different wavelength ranges fit (e.g.\ around the Fe-L or Fe-K$\alpha$ lines; \citealt{mazzotta:2004.xray.temperature.measurement.modeling.and.caveats,ghizzardi:2021.iron.cluster.profiles.sensitivity.of.fitting.different.lines.different.metallicities}); (c) different \textit{spatial} resolution (rebinning the same spectra in different annular intervals; \citealt{sanders:2002.centaurus.metallicity.profile.older.data.lower.res.still.some.drop.detected,sanders:2016.centaurus.metallicity.profiles.zdrop,panagoulia:2015.cluster.metal.profiles.with.drops}); or fits (d) using different temperature priors (fitting the temperature to different wavelength ranges, or using line-ratio temperatures; \citealt{matsushita:2002.m87.virgo.cluster.obs.metallicity.drop.temperature.fitting.challenges,ghizzardi:2021.iron.cluster.profiles.sensitivity.of.fitting.different.lines.different.metallicities,zhuhone:2023.cluster.temperature.fitting.sensitivities.simulations}); or (e) using single-versus-multi-temperature model fits \citep{mazzotta:2004.xray.temperature.measurement.modeling.and.caveats,vijayan:2022.cluster.cgm.multitemperature.fit.challenges}. This strongly influences the central abundances and inferred properties from the lines even in extremely well-observed clusters like Virgo, Centaurus, Perseus, and Ophiuchus (references above).  Accounting for pseudo-thermal continuum emission from CR-IC will only increase the degeneracy in this modeling.   
Resolving these challenges will require a combination of high spatial/angular resolution (few arcsecond or better, to resolve $\ll 10$\,kpc scales in nearby clusters, which is where we predict any actual signal relative to UV/optical), high spectral resolution microcalorimeter (few eV, to resolve the line ratios), with high sensitivity (for sufficient signal-to-noise without downgrading/rebinning the spatial/spectral resolution), of the sort only possible with future proposed missions like ATHENA (as compared to present studies with XRISM/Chandra/XMM/etc.).

\section{Conclusions}
\label{sec:conclude} 

Given their prodigious production of relativistic leptons, radio sources in clusters should produce inverse Compton emission halos  which have X-ray spectra resembling $\sim$\,keV thermal emission, X-ray surface brightness $\propto 1/R$, and extend to $\sim 100\,$kpc before being truncated by losses. 
These are the inevitable counterparts of the (more compact) radio minihalos seen in many bright radio clusters, and ubiquitous ultra-steep spectrum regions.  As radio-emitting leptons cool (age) out of radio bands (becoming undetectable) the X-ray continuum emission becomes increasingly thermal continuum-like, dominated by IC scattering of the thermal CMB by $\sim$ GeV electrons, with cooling ages $\sim$\,Gyr (Fig.~\ref{fig:spectrum}). In low-mass halos, this resembles ``hot'' (super-virial) gas at large radii in the halo, and (given known leptonic injection rates from Milky Way-like galaxies into the CGM), can explain the otherwise puzzling extended soft X-ray eROSITA halos observed \citep{zhang:2024.hot.cgm.around.lstar.galaxies.xray.surface.brightness.profiles}. In high-mass halos, this resembles ``cool'' (sub-virial) gas in cluster centers, i.e.\ cool cores (CCs); see Fig.~\ref{fig:mhalo}.  We show that for typical observed radio source strengths in strong CCs, the CR-IC halo can  dominate the X-ray continuum luminosity in the central $\lesssim 100\,$kpc, and produces a surface brightness and ``apparent'' (inferred from X-ray fitting, assuming only thermal emission) density, temperature, entropy and pressure profile all similar to those observed in SCCs (Fig.~\ref{fig:profiles}). 

If CRs contribute significantly to observed CC X-ray emission, this  provides a natural explanation for many observed correlations and puzzles. 
For example, the classical cooling flow problem -- that there is orders-of-magnitude less cold gas and star formation than predicted given the naive apparent X-ray cooling rate of the diffuse halo in CCs -- is immediately explained because the apparent ``cooling luminosity'' $L_{\rm X,\,cool}$ in CCs is boosted to much higher values by CR-IC:  the true hot gas densities are then significantly smaller than suggested by a pure thermal interpretation of the X-ray emission.    Likewise, there is an observed close correlation between the apparent X-ray ``cooling luminosity'' $L_{\rm X,\,cool}$ in CCs, and the radio luminosity and/or apparent ``cavity/jet power'' provided by the AGN \citep{rafferty:2006.cluster.cavity.jet.power.vs.xray.luminosity.agn.feedback.arguments.but.accretion.rates.dont.match.obvious}. This is naturally reproduced if CR-IC contributes significantly to $L_{\rm X,\,cool}$, because these all trace the same leptonic source (Fig.~\ref{fig:pjet.vs.lx}). In short, in the most extreme CR-IC scenario, the AGN powers the \textit{apparent} cooling flow, rather than the cooling flow powering the AGN.

In addition to explaining existing SCC observations and correlations, we argue that the CR-IC scenario makes two unique observable predictions. First, and most direct, if CR-IC contributes significantly to CC X-ray emission at a given radius $r$, then the pressure $P_{\rm X}(r)$ inferred from X-ray spectral fits will generally be biased high relative to the true thermal pressure $P_{\rm true,\,therm}(r)$. The latter would be correctly recovered by SZ measurements, however, so this predicts $P_{\rm SZ}/P_{\rm X}$ decreases from a value $\approx 1$ at $r \ll 100\,$kpc in SCC clusters, roughly as the square root of the fraction of surface brightness coming from true thermal emission. 
Indeed, there appears to be tentative evidence of this decline in SZ vs.\ X-ray pressure in existing arcsecond-resolution SZ observations of CCs \citep{romero:2017.cluster.pressure.profiles.highres.sz.xray.cool.cores.show.central.pressure.deficit,romero:2020.cluster.pressure.profile.sz.xr.zw3146.xr.only.large.radii.agrees.there}, which argue for $P_{\rm SZ}/P_{\rm X} \sim 0.3$ in the central $\lesssim 30$\,kpc of SCCs (i.e.\ $\sim 10\%$ of the emission being true thermal); see Fig.~\ref{fig:sz}. 
Second, since CR-IC contributes  continuum luminosity, the equivalent widths of lines in the X-ray spectra will be diluted, leading to apparently suppressed metallicities (relative to UV/optical diagnostics) in CC centers at $\lesssim 30\,$kpc or so. Modeling this quantitatively requires knowing the \textit{true} underlying multi-phase thermal gas density, temperature, and metallicity structure, plus how CRs modify line emission both via thermal (Coulomb+adiabatic+streaming) heating and direct+secondary ionization/excitation. And it is well-known that observational inferences of metallicity and line ratios in CCs are non-trivial.
However, the qualitative effect of dilution of the line emission by CR-IC continuum appears to be observed: in almost all nearby (resolved at $\ll 10\,$kpc, or $\sim 1^{\prime\prime}-5^{\prime\prime}$) SCC clusters, the inferred metallicity from single-temperature X-ray fits appears significantly lower than the metallicity (typically $\sim 1.5-2\,Z_{\odot}$) obtained from standard optical/UV metallicity estimators (of both stellar and gas phases) of the same galaxies at the same radii (Fig.~\ref{fig:Z}). 

The models of CR-IC presented in this paper assume for simplicity that the central AGN injects CR leptons at a fixed rate.   This produces the spectra shown in Figure \ref{fig:spectrum} in which the CR-IC emission is monotonically lower `temperature' at larger radii, since the CRs at larger radii have cooled/aged more.   In reality, the continuum emission produced by CR-IC could be more complex, e.g. if the AGN jet power and/or  the CR streaming speed vary significantly in time.   This could produce CR-IC spectra with multiple components analogous to multi-temperature thermal emission.   The hardness of the IC emission could also vary with angle at fixed radius since the CRs are likely preferentially injected along the jet axis.   Our models also focus on lepton injection and neglect protons, motivated by observations of compact radio and $\gamma$-ray emission in the radio galaxies of interest \citep{bottcher:2013.blazar.modeling.almost.all.blazars.better.fit.by.leptonic.cr.models.not.hadronic,blandford:2019.agn.jets.review,cerruti:2020.agn.jet.leptonic.hadronic.review}. We explore a wide range of hadron-to-lepton ratios in \papertwo, and show that an LISM value of $\dot E_{\rm cr,  p}/\dot E_{\rm cr, \ell} \sim 15-50$ would likely overpredict $\gamma$-ray emission via pion decay (in some extremely deeply-studied nearby clusters) and produce very large CR proton pressures in cluster cores, but a hadronic component comparable to or a few times larger than the leptonic one would not change our conclusions, and would have little effect on observables (still order-of-magnitude below state-of-the-art $\gamma$-ray limits). 

Our interpretation of CC emission as IC alleviates the cooling flow problem in the sense that it implies that the gas densities in CC cluster cores may be significantly lower than expected based on thermal models of the X-ray emission (and hence the cooling times are longer and the cooling rates are lower).  However, this of course does not imply that AGN feedback is unimportant. Indeed our interpretation implies that observations of CR-IC X-ray emission directly constrain AGN feedback in the form of CRs. Moreover, the energetic requirements for AGN in clusters -- that there is some feedback energy injected by the AGN comparable to the cooling luminosity of the hot halo $L_{X,\,{\rm cool}}$ -- is similar in the standard interpretation (feedback balancing $L_{X,\,{\rm cool}}$) and our proposed one (CRs sourcing apparent $L_{X,\,{\rm cool}}$). The key difference is that in the CR-IC interpretation
the classic problem of how AGN feedback can be efficiently and isotropically coupled to all of the cooling gas is less severe because the thermal gas density is lower and the gas cooling times are longer.   

There are many calculations and observational comparisons which remain for future work. In \papertwo, we consider more detailed modeling of multi-wavelength observations including central radio mini-halos (from $10^{6}-10^{10}$\,Hz), $\gamma$-rays, hard X-rays ($20-100\,$keV), and EUV, detailed X-ray spectral modeling, as well as a wide range of other observations of CC properties (CC radii, cavity power, radio power, AGN luminosities, cavity/bubble sizes, etc.), and their dependence on both CR injection properties and redshift. In the nearest, best-studied clusters like Perseus and Virgo, there are a wealth of spatially-and-spectrally-resolved observations that can be compared to constrain models like those here, so it will be particularly useful to consider detailed forward-modeling of specific observed systems.

Observations do unambiguously show that CC clusters contain much more multiphase gas than NCC clusters (e.g., \citealt{Cavagnolo2008}).  This implies that the thermal gas properties of CC clusters must be distinct  from those of NCC clusters.  Indeed, we argue (as in nearly all previous work) that this is the origin of the strong AGN at the center of CC clusters (Fig.~\ref{fig:cartoon}).   There is, however, ample room for the thermal gas density in CC clusters to be larger than that in NCC clusters but less than standard estimates based on the X-ray continuum emission.    Observations of the multiphase gas in CCs can be used to constrain non-thermal ionization by CRs, particularly molecular gas \citep{ferland:2009.particle.ionization.needed.for.molecular.line.emission.in.cc.perseus}; this is a promising route for better constraining the CR population in CC clusters.   Further ahead, future radio instruments with sensitivity and angular resolution similar to LOFAR but at much lower frequencies could potentially see the radio synchrotron counterparts of these X-ray halos (Fig.~\ref{fig:spectrum}). Hypothetical future $\gamma$-ray instruments may also detect their emission, but it would require $\sim 1000$ times the sensitivity of Fermi with $\lesssim 10''$ angular resolution, at energies $\sim 10-100\,$MeV. 

There are, of course, scenarios in which the CR-IC X-ray halos predicted here could be fainter than we anticipate.  One is if the GeV particles diffuse/stream at much higher speeds than assumed here; this would suppress the CR energy density and CR-IC surface brightness at a given radius relative to the models we have presented.   Alternatively if the GeV particles cannot escape strong magnetic field regions near the source (e.g., \citealt{Ewart2024}), most of their energy would come out in low frequency radio emission rather than IC X-ray emission.  More detailed comparison of the model developed here to observations will assess the role of CR-IC scattering in the X-ray emission of CC clusters.

\begin{acknowledgements}
We thank John ZuHone, Mark Voit, Iryna Zhuravleva, Peng Oh, and our anonymous referees for insightful comments and suggestions. Support for PFH was provided by a Simons Investigator Grant. This work began during PFH's sabbatical at the IAS.
\end{acknowledgements}

\bibliographystyle{mn2e}
%\bibliography{/Users/phopkins/Dropbox/Public/ms}
\bibliography{ms_extracted}

\end{document}